\documentclass[12pt,twocolumn,a4paper]{aastex631}
\hypersetup{linkcolor=cyan,citecolor=blue,urlcolor=cyan}

\usepackage{xspace}

\usepackage{float}
\usepackage[caption = false]{subfig}
\usepackage{xcolor}

\usepackage{xspace}
\newcommand{\Ha}{\ifmmode {\mathrm{H}\alpha} \else H$\alpha$\fi\xspace}
\newcommand{\Hb}{\ifmmode {\mathrm{H}\beta} \else H$\beta$\fi\xspace}

\newcommand{\Hii}{\ifmmode \rm{H}\,\textsc{ii} \else H~{\textsc{ii}}\fi\xspace}
\newcommand{\Hi}{\ifmmode \rm{H}\,\textsc{i} \else H~{\textsc{i}}\fi\xspace}
\newcommand{\Nii}{\ifmmode [\text{N}\,\textsc{ii}]\lambda 6584 \else [N~{\scshape ii}]$\lambda 6584$\fi\xspace}
\newcommand{\nii}{\ifmmode [\text{N}\,\textsc{ii}] \else [N~{\scshape ii}]\fi\xspace}
\newcommand{\Oii}{\ifmmode [\rm{O}\,\textsc{ii}]\lambda 3727 \else [O~{\textsc{ii}}]$\lambda$3727\fi}
\newcommand{\oii}{\ifmmode [\rm{O}\,\textsc{ii}] \else [O~{\textsc{ii}}]\fi}
\newcommand{\Oiii}{\ifmmode [\rm{O}\,\textsc{iii}]\lambda 5007 \else [O~{\textsc{iii}}]$\lambda$5007\fi}
\newcommand{\oiii}{\ifmmode [\rm{O}\,\textsc{iii}] \else [O~{\textsc{iii}}]\fi}

\shorttitle{Post-starburst galaxies in intermediate redshift clusters}
\shortauthors{A. Werle et al.}

\begin{document}

\title{Post-starburst galaxies in the centers of intermediate redshift clusters}

\correspondingauthor{Ariel Werle}
\email{ariel.werle@inaf.it}

\author[0000-0002-4382-8081]{Ariel Werle}
\affiliation{INAF- Osservatorio Astronomico di Padova, Vicolo Osservatorio 5, 35122 Padova, Italy}

\author[0000-0001-8751-8360]{Bianca Poggianti}
\affiliation{INAF- Osservatorio Astronomico di Padova, Vicolo Osservatorio 5, 35122 Padova, Italy}

\author[0000-0002-1688-482X]{Alessia Moretti}
\affiliation{INAF- Osservatorio Astronomico di Padova, Vicolo Osservatorio 5, 35122 Padova, Italy}

\author[0000-0002-6179-8007]{Callum Bellhouse}
\affiliation{INAF- Osservatorio Astronomico di Padova, Vicolo Osservatorio 5, 35122 Padova, Italy}

\author[0000-0003-0980-1499]{Benedetta Vulcani}
\affiliation{INAF- Osservatorio Astronomico di Padova, Vicolo Osservatorio 5, 35122 Padova, Italy}

\author[0000-0002-7296-9780]{Marco Gullieuszik}
\affiliation{INAF- Osservatorio Astronomico di Padova, Vicolo Osservatorio 5, 35122 Padova, Italy}

\author[0000-0002-3585-866X]{Mario Radovich}
\affiliation{INAF- Osservatorio Astronomico di Padova, Vicolo Osservatorio 5, 35122 Padova, Italy}

\author[0000-0002-7042-1965]{Jacopo Fritz}
\affiliation{Instituto de Radioastronom\'ia y Astrof\'isica, Universidad Nacional Aut\'onoma de M\'exico, Morelia, Michoac\'an, 58089 M\'exico\\}

\author[0000-0003-1581-0092]{Alessandro Ignesti}
\affiliation{INAF- Osservatorio Astronomico di Padova, Vicolo Osservatorio 5, 35122 Padova, Italy}

\author[0000-0001-5492-1049]{Johan Richard}
\affiliation{Univ Lyon, ENS de Lyon, CNRS, Centre de Recherche Astrophysique de Lyon UMR5574, 69230 Saint-Genis-Laval,France\\}

\author[0000-0001-9976-1237]{Geneviève Soucail}
\affiliation{Institut de Recherche en Astrophysique et Planétologie (IRAP), Université de Toulouse, CNRS, UPS, CNES,14 Av. Edouard Belin, 31400 Toulouse, France\\}

\author[0000-0002-6971-5755]{Gustavo Bruzual}
\affiliation{Instituto de Radioastronom\'ia y Astrof\'isica, Universidad Nacional Aut\'onoma de M\'exico, Morelia, Michoac\'an, 58089 M\'exico\\}

\author[0000-0003-3458-2275]{Stephane Charlot}
\affiliation{Sorbonne Universit\'es, UPMC-CNRS, UMR7095, Institut d'Astrophysique de Paris, F-75014, Paris, France\\}

\author[0000-0003-2589-762X]{Matilde Mingozzi}
\affiliation{Space Telescope Science Institute, 3700 San Martin Drive, Baltimore, MD 21218, USA}

\author[0000-0002-8372-3428]{Cecilia Bacchini}
\affiliation{INAF- Osservatorio Astronomico di Padova, Vicolo Osservatorio 5, 35122 Padova, Italy}

\author[0000-0002-8238-9210]{Neven Tomicic}
\affiliation{INAF- Osservatorio Astronomico di Padova, Vicolo Osservatorio 5, 35122 Padova, Italy}

\author[0000-0001-5303-6830]{Rory Smith}
\affiliation{Korea Astronomy and Space Science Institute (KASI), 776 Daedeokdae-ro, Yuseong-gu, Daejeon 34055, Republic of Korea}

\author[0000-0002-0431-2445]{Andrea Kulier}
\affiliation{INAF- Osservatorio Astronomico di Padova, Vicolo Osservatorio 5, 35122 Padova, Italy}

\author[0000-0001-5766-7154]{Giorgia Peluso}
\affiliation{INAF- Osservatorio Astronomico di Padova, Vicolo Osservatorio 5, 35122 Padova, Italy}
\affiliation{Dipartimento di Fisica e Astronomia ``G. Galilei'', Università di Padova, Vicolo dell’Osservatorio 3, 35122 Padova, Italy
}

\author[0000-0001-5766-7154]{Andrea Franchetto}
\affiliation{INAF- Osservatorio Astronomico di Padova, Vicolo Osservatorio 5, 35122 Padova, Italy}
\affiliation{Dipartimento di Fisica e Astronomia ``G. Galilei'', Università di Padova, Vicolo dell’Osservatorio 3, 35122 Padova, Italy
}

\begin{abstract}
We present results from MUSE spatially-resolved spectroscopy of 21 post-starburst galaxies in the centers of 8 clusters from $z\sim0.3$ to $z\sim0.4$.
We measure spatially resolved star-formation histories (SFHs), the time since quenching ($t_Q$) and the fraction of stellar mass assembled in the past 1.5 Gyr ($\mu_{1.5}$).
The SFHs display a clear enhancement of star-formation prior to quenching for 16 out of 21 objects, with at least 10\% (and up to $>50$\%) of the stellar mass being assembled in the past 1.5 Gyr and $t_Q$ ranging from less than 100 Myrs to $\sim800$ Myrs.
By mapping $t_Q$ and $\mu_{1.5}$, we analyze the quenching patterns of the galaxies. 
Most galaxies in our sample have quenched their star-formation from the outside-in or show a side-to-side/irregular pattern, both consistent with quenching by ram-pressure stripping. 
Only three objects show an inside-out quenching pattern, all of which are at the high-mass end of our sample. At least two of them currently host an active galactic nucleus. 
In two post-starbursts, we identify tails of ionized gas indicating that these objects had their gas stripped by ram pressure very recently.
Post-starburst features are also found in the stripped regions of galaxies undergoing ram-pressure stripping in the same clusters, confirming the link between these classes of objects.
Our results point to ram-pressure stripping as the main driver of fast quenching in these environments, with active galactic nuclei playing a role at high stellar masses.
\end{abstract}

\keywords{galaxies: clusters: general --- galaxies: general --- galaxies: evolution}

\section{Introduction} \label{sec:intro}

Post-starburst (also known as k+a or E+A) galaxies are a population of galaxies undergoing a rapid transition from star-forming to quiescent. 
These objects were first identified as a distinct population in clusters at intermediate redshift \citep{Dressler1983, Couch1987} but have since been recognized in a variety of environments and redshifts \citep[e.g][]{Poggianti2004}.
Post-starbursts are usually identified by their strong Balmer absorption lines and lack of emission lines associated with star-formation \citep[e.g][]{Poggianti1999, Dressler1999, Goto2007}, although recently other selection methods have been developed (e.g \citealt{Wild2007, Wild2014, Belli2019}; see \citealt{French2021} for a description of various selection criteria). 
These characteristic spectral features indicate significant levels of star-formation in the recent past ($\sim1-1.5$ Gyr) but no ongoing star-formation, indicating fast quenching sometime in the recent past.

There is strong evidence that post-starbursts were more common at higher redshift in all environments. In fact \cite{Deugenio2020b} has recently shown that the stacked spectra of 9 massive quiescent field galaxies at $z\sim3$ has clear post-starburst features.
\cite{Wild2009} finds that the number density of post-starbursts decreases by a factor of 200 from $z\sim0.7$ to $z\sim0.07$ and that they account for $38^{+4}_{-11}\%$ of the growth of the red sequence at $0.5<z<1.0$.
Post-starbursts represent less then $1\%$ of the field galaxy population at $z\sim0.5$ and more then $5\%$ at $z\sim2$ \citep{Wild2016}. 
Using a different approach, \cite{Belli2019} found that $34\%$ of quiescent galaxies are post-starbursts at $z\sim2.5$, while at $z\sim1$ they are only $4\%$. 
\cite{Fritz2014} found that the fraction of bright post-starbursts in cluster decreases from 18\% at redshifts $0.37<z<0.56$ to 4.6\% at $0.04<z<0.07$.  
The mass function of post-starburst galaxies also evolves with redshift in all environments, shifting towards lower masses at lower redshift according to a downsizing trend \citep{Poggianti2004, Wild2016}.

Post-starburst galaxies are also known to favor denser environments, being more common in clusters \citep{Poggianti2009, Dressler2013, Paccagnella2017, Paccagnella2019, Wilkinson2021}.
The main process responsible for fast quenching in these environments seems to be ram-pressure stripping.
For example, \cite{Poggianti2009} finds a correlation between the fraction of post-starbursts and the cluster velocity dispersion.
This correlation establishes a link between the rapid quenching of star-formation and the properties of the intra-cluster medium (ICM), as ram pressure stripping becomes more efficient in more massive clusters.
A perhaps more direct evidence of this connection is found by \cite{Poggianti2004}, where the authors identify a spatial correlation between post-starbursts and substructures of the hot ICM detected using X-Ray data from XMM-Newton.
When studying cluster galaxies in the GASP \citep[GAs Stripping Phenomena in galaxies][]{Poggianti2017} survey, \cite{Vulcani2020} finds that the star-formation histories and location within the cluster of rapidly quenched galaxies is consistent with quenching by ram-pressure stripping.
Furthermore, \cite{Marco2017} and \cite{Poggianti2019JW100} showed that regions of jellyfish galaxies (i.e galaxies with long tails of gas stripped by ram-pressure) that were already stripped of gas have spectral features typical of post-starburst galaxies, further suggesting that galaxies undergoing ram pressure stripping could be the progenitors of post-starbursts.

In the field and group environments, post-starbursts are usually associated with other processes, most commonly galaxy interactions and mergers.
Simulations show that gas-rich major mergers can lead to centrally concentrated starbursts, followed by a post-starburst phase and significant morphological transformation \citep{Wild2009, Bekki2005, Snyder2011, Pawlik2019, Zheng2020}. These features are consistent with several observational studies of field/group post-starbursts \citep{Blake2004, Pawlik2018, Chen2019, Deugenio2020a}.
Using spatially resolved spectroscopy from the MaNGA \citep[Mapping Nearby Galaxies at Apache Point Observatory][]{MaNGA} survey, \cite{Rowlands2018} found that more asymmetric galaxies have a larger fraction of post-starburst spaxels, which also favors the merger hypothesis.
Most post-starbursts in the field have a spheroidal morphology expected of merger remnants \citep{Blake2004, Pawlik2018}, while cluster post-starbursts have more prominent disks that indicates quenching through a process that does not significantly disturbs the structure of the stellar disk, such as ram-pressure stripping \citep{Dressler1999, Tran2003}.

Post-starbursts are also known to be connected with feedback from active galactic nuclei \citep[AGN,][]{Yan2006, Sanmartim2013, Pawlik2018}.
However, it is unclear if AGN are efficient enough to be the main driver of quenching \citep[e.g][]{Kaviraj2007PSB,Baron2018} or only prevent any residual star-formation from taking place after a main quenching event is triggered by a different process \citep[e.g][]{Yesuf2014}.

In the local universe, ``rejuvenation events'' in early-type galaxies \citep{Werle2020, Camila2021} triggered by minor mergers can also lead to a post-starburst phase \citep{Pawlik2018, Dressler2013}.
In particular, \cite{Pawlik2018} finds that this process is the origin of $\sim40$\% of massive post-starbursts at $z<0.05$. However, \cite{Chauke2019} finds rejuvenated galaxies at $z\sim0.8$ to have weaker Balmer lines than typical post-starbursts.

%Paccagnella2017 -> cluster PSBs
%Rowlands2018 -> fraction of PSB regions

Other  processes can also lead to the quenching of star-formation, but act on longer timescales that are inconsistent with the spectral features of post-starbursts. 
The heating or removal of gas in the circumgalactic medium \citep[CGM, see][]{Tumlinson2017} via ram-pressure or tidal interactions can prevent the accretion of new gas onto galaxy disks, a processes referred to as ``starvation'' \citep{Larson1980}. This leads to the cessation of star-formation once the remaining gas in the disk is consumed.
Since the gas-depletion times in normal star-forming galaxies are longer than 1 Gyr \citep{Saintonge2011,delosReyes2019}, this effect would lead to a very slow quenching process that cannot produce a post-starburst spectra.
We note, however, that the gas-depletion times for starburst galaxies are shorter than 1 Gyr \citep[see][]{Kennicutt2021}, indicating that the starvation of a starburst galaxy could in principle lead to a post-starburst. 
The cumulative effect of weak tidal interactions \citep[``harassement'',][]{Moore1996} also acts on long time scales and cannot lead to a post-starburst phase.

The extent to which each of these processes contributes to the quenching of star-formation in galaxies residing in different environments remains as an open topic.
In this work, we take advantage of MUSE (Multi Unit Spectroscopic Explorer) data collected by the GTO program to study a sample of post-starburst galaxies using rest-frame optical spatially resolved spectroscopy of the central regions of intermediate redshift clusters ($0.308<z<0.451$). This allows us to study how fast-quenching happens in different regions of galaxies in dense environments, as has been previously done for field and group galaxies \citep[e.g][]{Rowlands2018, Chen2019, Deugenio2020a}.

The paper is organized as follows: Section \ref{sec:data} describes the data-set and the sample selection. Our spectral synthesis method is described in section \ref{sec:synthesis} and results from the synthesis are presented in \ref{sec:results}. In section \ref{sec:special} we present some special cases of post-starbursts with distinct features in their emission line maps. Further clues on ram-pressure stripping are presented in \ref{sec:discussion} and conclusions are laid out in \ref{sec:conclusions}.

We assume a standard $\Lambda$CDM cosmology with $\Omega_{\rm M}=0.3$, $\Omega_\Lambda=0.7$ and $h=0.7$. We adopt the solar metallicity value of $Z_\odot=0.017$ and a \cite{Chabrier2003} initial mass function. The chosen epoch for right ascension and declination (RA and Dec) is J2000.

\section{Data and Sample}\label{sec:data}

\subsection{The data set}\label{sec:dataset}

This work is based on MUSE data cubes from the MUSE Lensing Cluster GTO program, which provides deep observations of the central regions of clusters. These fields were originally targeted due to the abundance of gravitationally lensed background sources. 
The pixel size in the redshift range of our work is of $\sim1$ kpc, while the seeing is of of $1^{\prime\prime}$ ($\sim4-5$ kpc). 
The cluster sample is extracted from a multitude of surveys, including the MAssive Clusters Survey
\citep[MACS,][]{MACS}, the Frontier Fields program \citep[FFs,][]{FFs}, the Grism Lens-Amplified Survey from Space
\citep[GLASS,][]{GLASS} and the Cluster Lensing And Supernova survey with Hubble \citep[CLASH,][]{CLASH}.
The complete data set is extensively described by \cite{Richard2021}.

Here we use data from the clusters Abell 2744, Abell 370, MACS J1206.2-0847, MACS J0257.6-2209,  RX J1347.5-1145, SMACS J2031.8-4036, SMACS J2131.1-4019 and Abell S1063.
Basic information about these clusters is given in table \ref{table:clusters}, where we include the central redshift, position, and the velocity dispersion of galaxies in the cluster ($\sigma$).
Redshifts, positions and $\sigma$ were extracted from \cite{Richard2021}, except for AS1063 for which these data come from \cite{Sartoris2020} (position and redshift) and \cite{Gomez2012} ($\sigma$). The cluster sample is quite massive as it was selected to study gravitational lensing. This characteristic also makes the sample ideal for finding post-starburst galaxies, as the frequency of post-starburst galaxies is known to increase with cluster mass \citep{Poggianti2009, Dressler2013, Paccagnella2019}.

% The data shown in the table was extracted from \cite{Owers2011} for Abell 2744, \cite{Lah2009} for Abell 370, \cite{Zitrin2012} for the BCG position and \cite{Ebeling2009} for redshift and $\sigma$ of MACS J1206.2-0847, 

\begin{table*}[]
\centering
\caption{Basic information about the clusters studied in this work.}
\begin{tabular}{llllll}\label{table:clusters}
Cluster                  & Redshift    & RA  & Dec     & $\sigma\mathrm{[km/s]}$ \\
\hline
Abell 2744               & 0.308    & 00:14:20.702 & -30:24:00.63   &  1357 \\
Abell 370                & 0.375    & 02:39:53.122 & -01:34:56.14   &  1789 \\
MACS J1206.2-0847        & 0.438    & 12:06:12.149 & -08:48:03.37   &  1842 \\
MACS J0257.6-2209        & 0.322    & 02:57:41.070 & -22:09:17.70   &  1633  \\
RX J1347.5-1145          & 0.451    & 13:47:30.617 & -11:45:09.51   &  1097 \\
SMACS J2031.8-4036       & 0.331    & 20:31:53.256 & -40:37:30.79   &  1531 \\
SMACS J2131.1-4019       & 0.442    & 21:31:04.831 & -40:19:20.92   &  1378 \\
Abell S1063          & 0.3458   & 22:48:43.99  & -44:31:50.98   &  1660 \\
\hline
\end{tabular}
\end{table*}

For the clusters MACS J0257.6-2209, RX J1347.5-1145 and SMACS J2131.1-4019, observations were carried out using the MUSE adaptive optics system. In these cases the region around $\sim5900$\AA\ in the observed-frame cannot be used for science due to the NaD notch filter \citep{MUSEAO}. 

In addition to the MUSE cubes, we also use images from the Hubble Space Telescope (HST) in the F814W filter, but these are shown only for illustrative purposes in sections \ref{sec:sample} and \ref{sec:tails}. 
These data are taken from the HST archive, for MACS fields, the images are re-reduced as presented in \cite{Richard2021}.

\subsection{Post-starburst galaxy sample}\label{sec:sample}

As described by \cite{Moretti2021}, each datacube was visually inspected to find post-starbursts, as well as the stripped galaxies presented in that paper. Post-starburst galaxies are identified as those generally lacking emission lines but still presenting strong Balmer absorption features anywhere in the stellar disk. After visual inspection, the rest-frame $H\delta$ and $H\beta$ equivalent widths ($H\delta_A$ and $H\beta_A$, respectively) were quantified to confirm the galaxies as post-starburst. We calculate $H\delta_A$ and $H\beta_A$ using the rest frame spectral regions of 4076-4088\AA\ and 4806-4826\AA\ for the blue continuum, 4091-4112\AA\ and 4826-4896\AA\ for the lines, and 4117-4136\AA\ and 4896-4918\AA\ for the red continuum.

In Table \ref{table:sample} we list some basic information about the 21 post-starbursts in our sample, namely the IDs, cluster names, redshifts, coordinates, $H\delta_A$ and $H\beta_A$ of the integrated spectra and stellar masses determined using our spectral synthesis procedure (see section \ref{sec:synthesis}). 
The IDs represent an abbreviation of the cluster name followed by an integer, for Abell S1063 we also include the acronyms NE (north east) and SW (south west) to indicate two different MUSE pointings.
% \bp{Do we really want to give the masses with two decimal cifers since the typical uncertainty is 0.3dex? Maybe one is sufficient? but then we have to change it throughout the paper. The same for the EWs....can we associate somehow an error for the values listed in Table 1? For the redshift instead in table 1 I would give 4 decimal cifers, 3 are too little.} 

\begin{table*}[]
\centering
\caption{Basic information about our sample of post-starburst galaxies. Equivalent widths marked with * were obtained from synthetic spectra.}
\begin{tabular}{llllllll}\label{table:sample}
ID           & Cluster                  & Redshift & RA & Dec & $H\delta_A\,\mathrm{[\AA]}$ & $H\beta_A\,\mathrm{[\AA]}$ & $\log\,M_\star/M_\odot$\\
\hline
A2744-01     & Abell 2744               & 0.291   &    00:14:19.754 & -30:23:57.83   &   6.8                 &  8.9    &    8.9  \\
A2744-02     & Abell 2744               & 0.320   &    00:14:20.175 & -30:23:56.77   &   4.0                 &  8.5    &    9.9    \\
A2744-05     & Abell 2744               & 0.307   &    00:14:18.806 & -30:23:13.48   &   6.2                 &  9.7    &    9.5    \\
A2744-07     & Abell 2744               & 0.299   &    00:14:20.474 & -30:23:15.10   &   4.2                 &  8.5    &    10.2   \\
A2744-08     & Abell 2744               & 0.305   &    00:14:18.975 & -30:24:00.287  &   4.8                 &  8.0    &    9.3   \\
A370-04      & Abell 370                & 0.361   &    02:39:51.370 & -1.33.59.091   &   7.5*                & 10.4    &    9.5   \\
A370-05      & Abell 370                & 0.390   &    02:39:53.544 & -01:34:31:751  &   5.3                 & 10.1    &    10.0    \\
MACS1206-06  & MACS J1206.2-0847        & 0.422   &    12:06:13.17  & -08:47:45.05   &   5.6                 &  9.5    &    10.7   \\
MACS1206-09  & MACS J1206.2-0847        & 0.427   &    12:06:11.36  & -08:48:22.0    &   4.9                 & 11.1    &    10.4   \\
MACS1206-11  & MACS J1206.2-0847        & 0.427   &    12:06:14.856 & -08:48:15.6    &   5.2                 &  9.2    &    10.5   \\
MACS0257-01  & MACS J0257.6-2209        & 0.335   &    02:57:42.581 & -22:09:20.61   &   6.6                 & 10.4    &    9.2   \\
MACS0257-03  & MACS J0257.6-2209        & 0.325   &    02:57:40.15  & -22:08:54.4    &   6.3                 &  8.7    &    9.5   \\
MACS0257-04  & MACS J0257.6-2209        & 0.323   &    02:57:40.7   & -22:09:21.6    &   5.4                 &  7.3    &    9.6   \\
MACS0257-05  & MACS J0257.6-2209        & 0.332   &    02:57:39.9   & -22:09:14.2    &   4.2*                &  7.8    &    9.9    \\
RXJ1347-02   & RX J1347.5-1145          & 0.431   &    13:47:31.36  & -11:45:50.7    &   8.5*                &  11.1   &    9.9    \\
SMACS2031-02 & SMACS J2031.8-4036       & 0.329   &    20:31:51.1   & -40:37:21.6    &   5.0                 &  8.5    &    10.1   \\
SMACS2131-05 & SMACS J2131.1-4019       & 0.459   &    21:31:03.0   & -40:19:06.2    &   4.9*                &  8.0    &    9.9    \\
SMACS2131-06 & SMACS J2131.1-4019       & 0.445   &    21:31:03.0   & -40:19:00.8    &   4.7*                &  8.5    &    9.5   \\
AS1063NE-01  & Abell S1063              & 0.332   &    22:48:44.22  & -44:31:41.0    &   5.1                 &  7.6    &    10.7   \\
AS1063NE-02  & Abell S1063              & 0.339   &    22:48:43.11  & -44:31:22.98   &   7.4                 & 10.3    &    9.8    \\
AS1063SW-01  & Abell S1063              & 0.326   &    22:48:40.11  & -44:32:04.6    &   7.4                 & 10.9    &    9.6   \\
\hline
\end{tabular}
\end{table*}

All visually inspected galaxies were confirmed as post-starbursts by comparison with typical criteria from the literature.
For most of our sample (14 galaxies), the integrated spectra are above the threshold of $H\delta_A=5$\AA, which is commonly used to select post-starbursts \citep[e.g][]{Goto2007, Alatalo2016}, others have slightly lower $H\delta_A$, but are also post-starbursts (k+a) according to the criteria of \cite{Poggianti1999} ($3<H\delta_A<5$\AA).
Note that for MACS0257-05, RXJ1347-02, SMACS2131-06, part of the spectral window required for calculating $H\delta_A$ is unavailable as it is affected by the NaD notch filter. In the case of A370-04, the 5577\AA\ sky line overlaps with the $H\delta$ region which brings down the $H\delta_A$ value and makes it unreliable.
Nevertheless, for these objects we report in the table the $H\delta_A$ values measured from the integrated model spectra obtained by SINOPSIS (see section \ref{sec:synthesis}); these are marked with * on the table.
Although $H\delta_A$ is the most commonly used index to select post-starbursts, other Balmer lines can also be used.
In the cases where $H\delta$ is missing, we confirm the post-starburst nature of the objects by their $H\beta_A$ index ($H\beta_A>2.5$\AA), as done by \cite{Vulcani2020} for local cluster galaxies. 

The galaxies are distributed in 8 clusters and span a redshift range from 0.291 for A2744-01 to 0.459 for SMACS2131-05. The ages of the universe in these limiting redshifts are 10.1 Gyr and 8.8 Gyr, respectively. Our sample spans a wide range of stellar masses, from $10^{8.9}M_\odot$ for A2744-01 to $10^{10.7}M_\odot$ for MACS1206-06. We note that our sample typically includes galaxies of low mass, mostly less massive than, e.g, the Milky Way. The low mass indicates that these galaxies could be very quickly stripped by ram-pressure in timescales shorter than the cluster crossing times, e.g before or soon after the pericenter passage.

\begin{figure}[ht]
\includegraphics[width=\columnwidth]{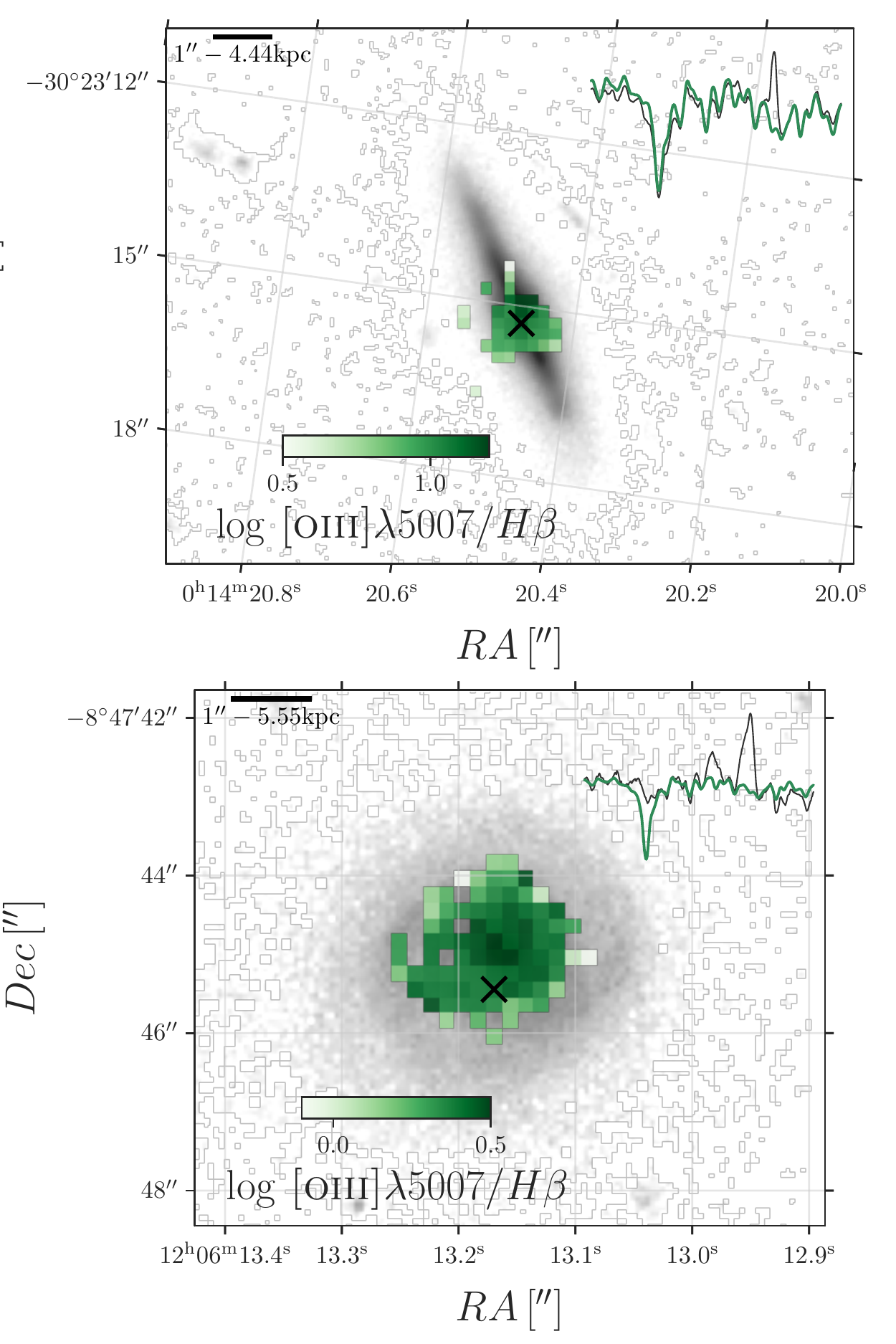}
   \caption{Maps of the $\Oiii / \Hb$ line ratio for two AGN candidates: A2744-07 (top) and MACS1206-06 (bottom). Only spaxels with $S/N>3$ in both $\Oiii$ and  $\Hb$ are included.
    HST F814W images are shown in gray scale in the background. On the top left we show the spectrum of a specific spaxel (marked with a dark x in the figure) around the $\Hb$ and $\Oiii$ lines (4776-5092\AA). The observed spectrum is plotted in black and the model obtained for the stellar continuum (see sec \ref{sec:synthesis}) is shown in green, both spectra are convolved with a 5 pixel box filter for clarity.}
\label{fig:agn_maps}
\end{figure}

\begin{figure}[ht]
\includegraphics[width=\columnwidth]{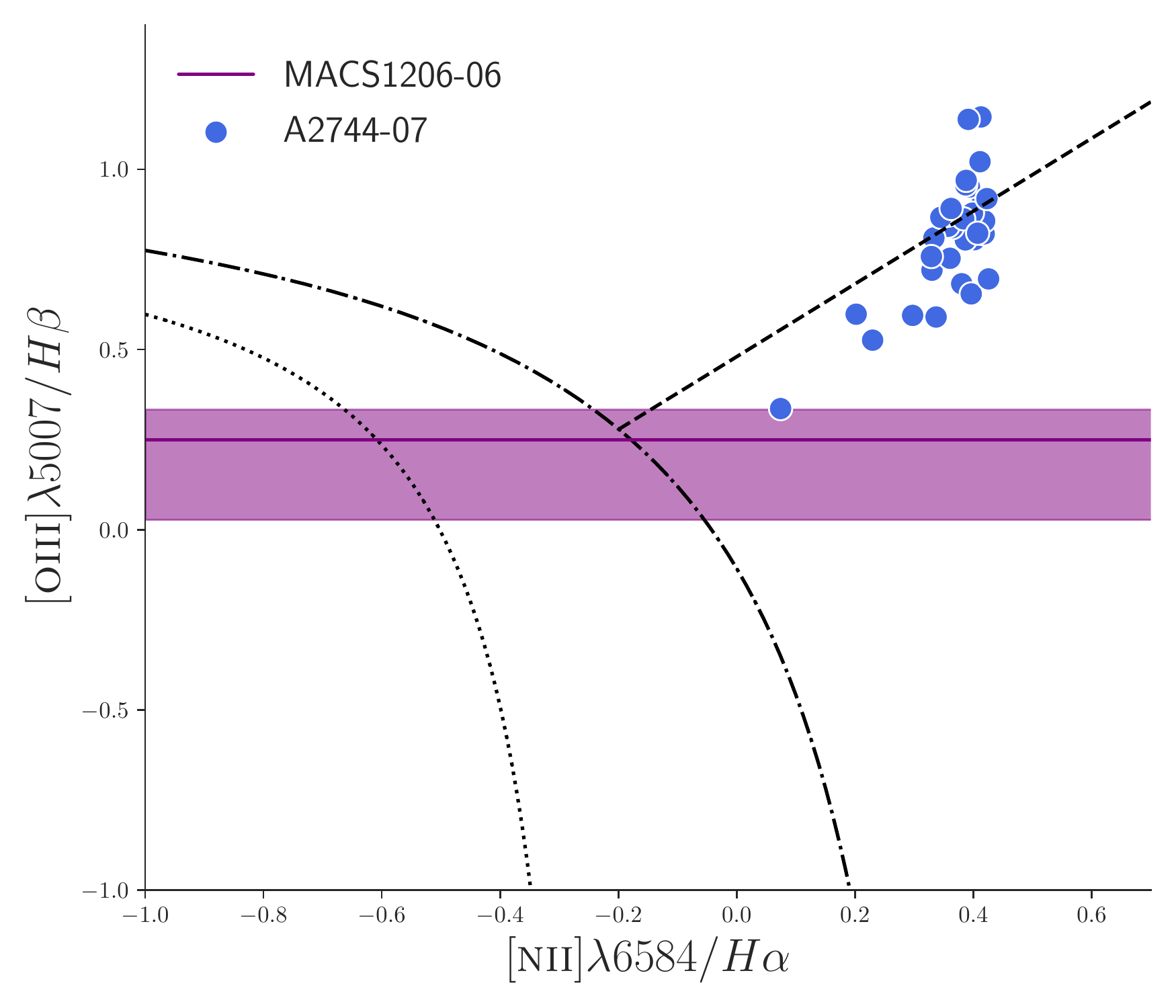}
   \caption{$\Nii/\Ha$ vs $\Oiii/\Hb$ BPT diagram. Blue points are spaxels from A2744-07 with $S/N>3$ in all lines in the diagram. The purple horizontal line is the median $\log \, \Oiii/\Hb$ for spaxels in MACS1206-06 with $S/N>3$ in $\Oiii$ and $\Hb$, the purple band indicates the region between the 10 and 90\% percentiles of $\Oiii/\Hb$ for the same spaxels.
   For reference, we plot the demarcation lines of \citet[dotted line]{Grazyna2006}, \citet[dot-dashed line]{Kewley2001} and the transposition of the \cite{Kewley2006} line proposed by \citet[dashed line]{Cid2010},  separating LINERs and Seyferts.}
\label{fig:bpt}
\end{figure}

We have included in our sample two objects that show emission lines in the central regions of their stellar disk, these are MACS1206-06 and A2744-07.
We include these objects because we interpret this emission as due to AGN. 
In Fig. \ref{fig:agn_maps}, we show maps of the $\Oiii/\Hb$ line ratio for these galaxies, including spaxels with signal-to-noise ratio ($S/N$) larger than 3 in both emission lines. 
The emission lines used here and throughout the paper were measured using the \textsc{highelf} code as described in \cite{Moretti2021}.
In the case of A2744-07 we are also able to measure the $\Ha$ and $\Nii/\Ha$ lines, allowing the full characterization of the central spaxels in a BPT diagram \citep{BPT}. Unfortunately, these lines are redshifted out of the MUSE spectral range for MACS1206-06. In Fig. \ref{fig:bpt} we analyze the central spaxels of these two objects in the $\Nii/\Ha$ vs $\Oiii/\Hb$ BPT diagram. For A2744-07 (blue points) we include spaxels with $S/N>3$ in all four lines. Since for MACS1206-06 we have only information on $\Oiii$ and $H\beta$, we can only constrain the y-axis of the diagram. In this case, we plot a horizontal line (purple line) indicating the median $\Oiii/\Hb$ and a band indicating the region between the 10 and 90\% percentiles of $\Oiii/\Hb$ in this galaxy, again considering only spaxels with $S/N>3$ in the two lines.

All spaxels in A2744-07 that satisfy the $S/N$ criteria lie above the \cite{Kewley2001} line (dot-dashed line in Fig. \ref{fig:bpt}), and can thus be considered as ionized by AGN.
% \mr{Actually they lie on the LINER-AGN boundary, which is not so uncommon for post-starburst galaxies. For instance, see Decker French et al. 2015, ApJ, 801, 1 (Fig.1, and sect. 2.5). Though it may be due to an AGN, other processes can't be excluded (shocks, post-AGB): if they are more unlikely than the AGN, it should be explained why (the same applies below).}
% The emission is centrally concentrated.
The BPT classification for MACS1206-06 is ambiguous: not only are we unable to constrain the x-axis, but the y-axis values fall in a region where star-formation regions and LINERs (which can be AGN) coexist. However, since an X-ray point source is detected within the galaxy in the Chandra catalog \citep{Ehlert2015, Wang2016} 
and there is no evidence for current star-formation from the spectral synthesis, we conclude that an AGN is the most likely ionization source. 

% \bp{Do we have an X-ray luminosity from Wang??} \mr{I agree that more details here would be very useful !}
% (iii) the high stellar mass of the galaxy high stellar mass of MACS1206-06 ($\log \, M_\star/M_\odot = 10.9$, see \citealt{Juneau2011})

For A2744-01 and SMACS2131-06 we also identify emission lines, but outside of the stellar disk. This will be discussed in section \ref{sec:tails}.

\section{Spectral Synthesis}\label{sec:synthesis}

In this section we will go into details regarding the spectral synthesis method and the determination of physical parameters for our sample of 21 post-starburst galaxies.
The spectral synthesis code is described in section \ref{sec:sinopsis}, details on the synthesis procedure are presented in section \ref{sec:procedure}, and in section \ref{sec:definitions} we present the definition of parameters calculated from the star-formation histories.

\subsection{Spectral synthesis code: SINOPSIS}\label{sec:sinopsis}

This work is heavily based on results from the spectral synthesis code SINOPSIS \citep[SImulatiNg OPtical Spectra wIth Stellar populations models][]{Fritz2007, Fritz2011, Fritz2014, Fritz2017}. SINOPSIS performs a non-parametric decomposition of galaxy spectra into a combination of stellar population models. This non-parametric approach allows the code to capture even very complex star-formation histories (\citealt{Leja2019, Conroy2013}, but see also \citealt{Cid2007}).

There are a few design choices that set SINOPSIS apart from other widely used non-parametric synthesis codes. First, unlike codes such as {\sc STARLIGHT} \citep{Cid2005} and {\sc pPXF} \citep[Penalized PiXel-Fitting][]{ppxf}, SINOPSIS does not perform full spectral fitting. Instead, the fitting procedure is performed in a selection of continuum band fluxes and spectral indexes (equivalent widths and $D_n4000$), which focuses the fitting efforts in parts of the spectrum that are more information-rich while also making the code less susceptible to problems regarding sky lines and other observational issues.
Another distinct design choice in SINOPSIS is the inclusion of emission features in the fitting of stellar population parameters. The emission lines produced by each stellar population element up to the age of 20 Myrs are modeled using {\sc cloudy} \citep{Cloudy}, allowing us to get better constraints for young components.
Finally, SINOPSIS allows for a full treatment of dust attenuation \citep{Charlot2000}. Each stellar population component can be fitted with an independent value of dust optical depth, although we do not take advantage of this feature in this work since our galaxies do not have young stars and this effect is not prominent.

In particular, including emission lines in the fit is very useful for modeling post-starburst galaxies. Since these galaxies do not have emission lines, this provides clear information to the code about the lack of current star-formation and prevents solutions that might include small fractions of young stellar populations that are poorly constrained from the optical stellar continuum \citep[see][]{Werle2019}.
The fitting of emission lines coupled with the versatility of a non-parametric method make SINOPSIS the most suitable tool for the current work.

\subsection{Synthesis ingredients and fitting procedure}\label{sec:procedure}

Our spectral synthesis method is based on stellar population spectra from the latest update of the \cite{Bruzual2003} models (Charlot \& Bruzual, in prep). In the optical region, these models rely on the MILES \citep{MILES} and IndoUS \citep{IndoUS} libraries of stellar spectra, which are combined according to the PARSEC isochrones from \cite{Chen2015} and \cite{PARSEC}. 
The spectral resolution of the models in the optical ($\sim2.50\mathrm{\AA}$ FWHM) is very similar to the one of MUSE spectra ($\sim2.55\mathrm{\AA}$ FWHM). 
In this work, we use models generated with a \cite{Chabrier2003} initial mass function.
A more detailed description of these models is provided in \cite{Werle2019}.

To generate the final set of model spectra used in our fitting procedure,
we combined the simple stellar population models from Charlot \& Bruzual into 16 age bins from 0 to 10 Gyrs\footnote{The limiting ages of the bins are (in years): $0$, $1.99\times10^{6}$, $3.98\times10^{6}$, $6.91\times10^{6}$, $1.99\times10^{7}$, $5.71\times10^{7}$, $2.028\times10^{8}$, $5.093\times10^{8}$, $8.072\times10^{8}$, $1.014\times10^{9}$, $1.435\times10^{9}$, $2.0\times10^{9}$, $3.0\times10^{9}$, $4.5\times10^{9}$, $6.25\times10^{9}$, $8.0\times10^{9}$ and $10\times10^{9}$.} and generated composite stellar populations assuming a constant SFR within the bins.

We use three values of metallicity: 0.004, 0.017 (solar) and 0.04.
The fitting is done separately for sets of models of different metallicities and the one with the lowest $\chi^2$ is chosen, i.e there is no chemical evolution in the derived SFHs.

The effects of dust 
can be modeled using 
independent values of extinction with a \cite{CCM} extinction law. However, since usually higher extinction values affect only stars younger than 20 Myrs and these are not identified in our galaxies due to the lack of emission lines, we use one single value of extinction.

In the fits used in this work, we include the equivalent widths of $H\alpha$, $H\beta$, $H\delta$ and calcium H and K lines, as well as the flux in 32 continuum bands. We note that the features used for each specific galaxy vary according to redshift. 

For each of our datacubes, we run SINOPSIS on all spaxels in a contiguous area around the center of the target galaxy where a stellar continuum can be detected, throughout this paper we refer to these as ``valid spaxels''. 
The total number of valid spaxels in all 21 galaxies in our sample is 2230, ranging from 36 for MACS0257-05 to 318 for MACS1206-06 with an average of 106 per galaxy. 
The redshifts for each spaxel, which are required for running SINOPSIS, were determined using {\sc pPXF} \citep{ppxf} on Voronoi binned regions with $S/N>5$ (as thoroughly described in \cite{Moretti2021}). For A2744-05, AS1063SW-01 and SMACS2131-06, foreground objects were manually masked out. 

To obtain the total stellar masses, we sum up the masses obtained with SINOPSIS for all valid spaxels and multiply it by a scale factor so the total mass corresponds to the entire region where the galaxy emission in the MUSE g-band image is 3$\sigma$ above the background level. 

\begin{figure}[ht]
\center
\includegraphics[width=\columnwidth]{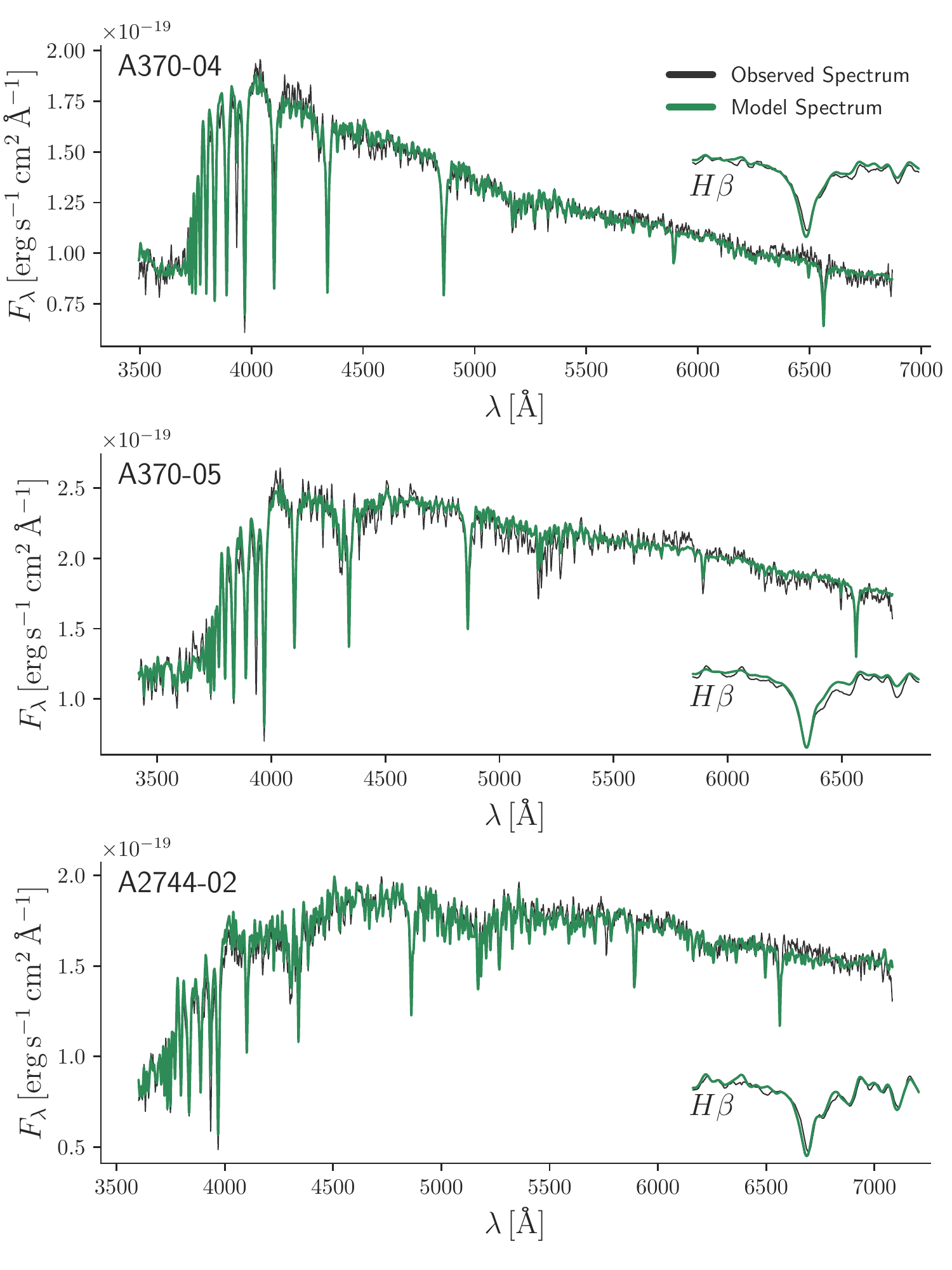}
   \caption{Three examples of SINOPSIS fits for post-starburst galaxies in our sample. Panels show integrated observed (gray) and model (green) spectra for (top to bottom) A370-04, A370-05 and A2744-02. 
   Spectra are convolved with a 5-pixel box filter to improve visualization.
   We include a zoom-in to the spectral region around the $H\beta$ line ($4861\pm75$\AA).}
\label{fig:fits}
\end{figure}

Fig. \ref{fig:fits} shows examples of SINOPSIS fits for individual spaxels in three post-starburst galaxies in our sample. The examples (A370-04, A370-05 and A2744-02) were chosen to illustrate different levels of post-starburst features.

\subsection{Star-formation history feature extraction}\label{sec:definitions}

To interpret the 2230 derived SFHs (one for each valid spaxel), we chose to reduce their dimensionality by extracting two relevant features: (i) The time since quenching ($t_Q$), and (ii) the fraction of stellar mass formed in the past 1.5 Gyr ($\mu_{1.5}$).
%and (iii) the enhancement of star-formation in the past 1.5 Gyr ($\epsilon$).

We define the time since quenching $t_Q$ as the time elapsed since the moment when the 
galaxy reaches 98\% of the total mass ever assembled, i.e counting also stars that have already died.
This parameter is calculated from a cumulative SFH obtained by converting the SFR of each age bin to mass fractions, summing the mass fractions cumulatively from the oldest to the youngest and linearly interpolating between points. We then find the age where the mass fraction is 98\%.
We choose not to use the 100\% mass fraction to avoid the influence of possible tiny insignificant fluctuations in the SFR. The value of 98\% is close enough to 100\% to give a robust definition of quenching without being affected by these fluctuations.

From the same cumulative SFHs we calculate the fraction of stellar mass assembled in the past 1.5 Gyr ($\mu_{1.5}$). 
The same quantity was explored also by \cite{Pawlik2018} and \cite{Wild2020}.
Although 1 Gyr would be closer to the lifetime of A-type stars that characterize the spectra of post-starburst galaxies, there can be a SFR enhancement starting slightly before that time.  
Thus, we chose the 1.5 Gyr limit as it includes the whole timescale over which the SFR may have been enhanced.
Nevertheless, the mass fractions calculated using 1 or 1.5 Gyr are very similar and this choice does not affect our results.
Note that here and throughout the paper we are not using a strict terminology when saying ``in the past 1.5 Gyr" or ``time since quenching" as we are referring to lookback times that have as reference the age of the universe in the galaxy redshift, and not at $z=0$ as would be implied by taking these sentences literally.

\section{Star-formation histories}\label{sec:results}

We now focus on the star-formation histories derived by SINOPSIS. In section \ref{sec:intsfh} we show the integrated SFHs of all galaxies, we then look at the distributions and maps of $t_Q$ and $\mu_{1.5}$ (sections \ref{sec:dists} and \ref{sec:maps}, respectively) and present our interpretation of the general trends in section \ref{sec:maptrends}.

\subsection{Integrated star-formation histories}\label{sec:intsfh}

\begin{figure*}[ht!]
\includegraphics[width=\textwidth]{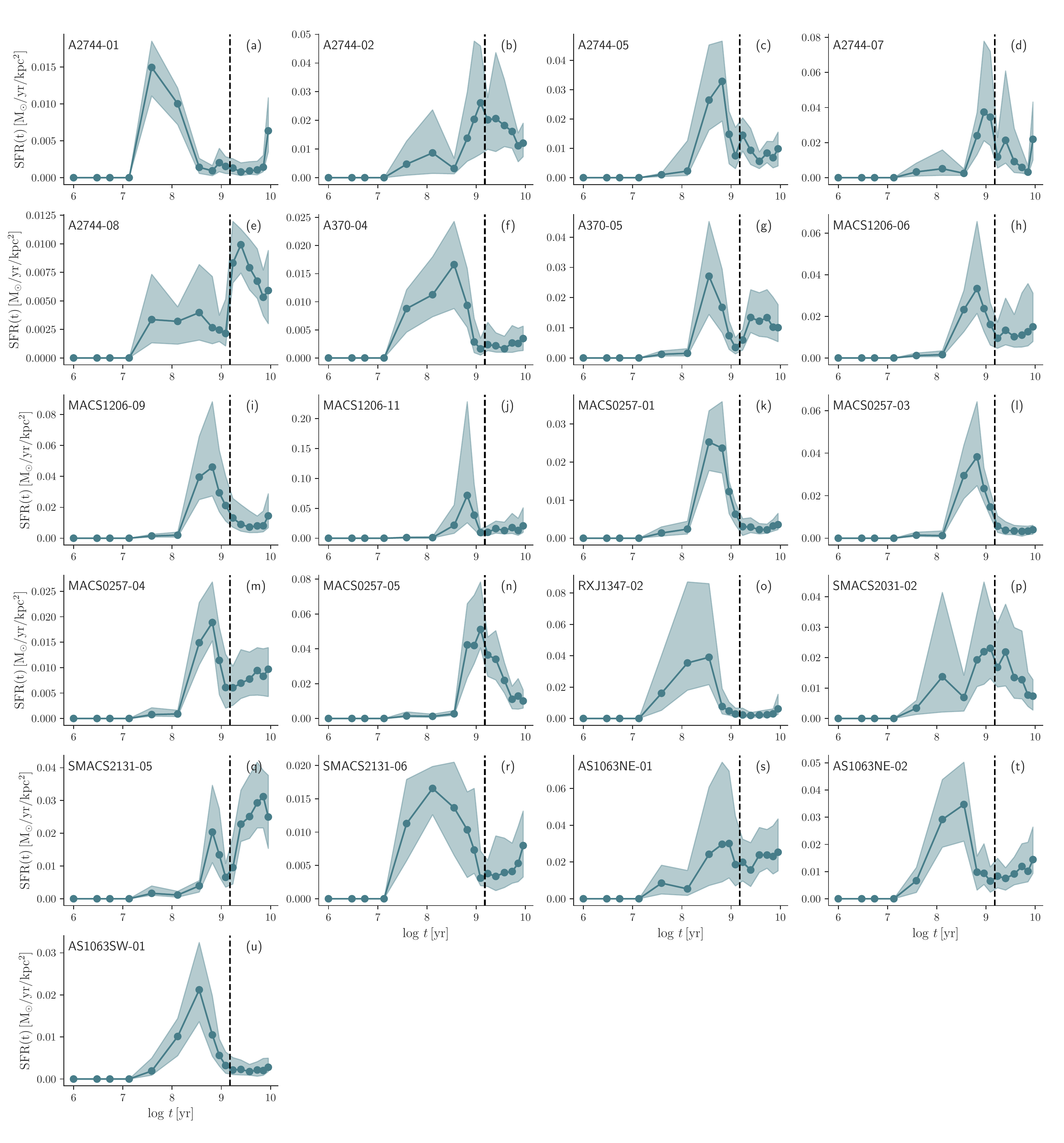}
   \caption{Median SFHs for galaxies in our sample, with shaded bans indicating the interquartile regions. 
   Points are plotted at the central lookbacktime of each age bin. 
   Dashed vertical lines indicate the reference lookback time of 1.5 Gyr. }
\label{fig:intsfh}
\end{figure*}

The definition of post-starburst is purely empirical and can accommodate a variety of SFHs. The lack of emission lines implies that $\mathrm{SFR(t<\;20Myr)}=0$, and the strong Balmer absorption lines indicate the presence of A-type stars whose lifetime is $\sim1$ Gyr. This constrains the galaxies to SFHs that were rapidly quenched in $t<1$ Gyr but does not indicate if the SFR was actually enhanced before quenching (i.e if there was a burst) or if the SFR suddenly dropped some time in the past 1 Gyr without any prior enhancement. 
To examine this, we have calculated the median of the spaxel-by-spaxel SFHs
for each galaxy, 
as shown in Fig. \ref{fig:intsfh}.  Median SFHs are plotted in blue with shaded regions indicating the interquartile regions (between the 25 and 75\% percentiles), a dashed line indicates a lookback time of 1.5 Gyr.
Note that we choose to represent the SFHs as SFR vs. $\log \, t$ to highlight the bursts, but we should keep in mind that older bins are much wider and thus low SFRs at old ages can still correspond to a very large fraction of mass formed.

For most galaxies (16 out of 21) it is possible to identify a clear and isolated burst of star-formation prior to quenching, and in some cases this burst is the main feature in the SFH. 
In other cases the SFHs are more complex and it is hard to identify a single burst. In the case of A2744-02, A2744-07 and SMACS2031-02 (panels b, d and p) there seems to be an enhancement in SFR close to the 1.5 Gyr mark followed by a smaller enhancement in more recent times. For A2744-08 (panel e) there is a strong drop in the SFR at around 1.5 Gyr ago after which SFR is sustained at an almost constant level until quenched. In the case of MACS0257-05 (panel n) there seems to be a burst 1.5 Gyr ago but the feature is unclear as the SFR was rising prior to that event.

Notwithstanding these peculiarities, all of these SFHs can be described as rapidly quenched, as all of them display a sharp decrease in the SFR at $t<$1 Gyr. Although in some cases (e.g. A2744-07, A2744-08, MACS0257-04 and SMACS2131-05; panels d, e, m and q) some residual star-formation is sustained after the main quenching event. 
This residual star-formation is associated with very small mass fractions that do not count towards our definition of quenched. That being said, 
it is important to have this effect in mind when analyzing the results for these galaxies. 

In most cases, the interquartile regions traced by the shaded bands in Fig. \ref{fig:intsfh} do not deviate from the median shape of the SFH, indicating that the SFH is somewhat similar across the spaxels of a single galaxy. Indeed, galaxies in our sample are generally post-starburst in all spaxels. Only 3.5\% of the spaxels for which $H\delta_A$ can be measured (91\% of the spaxels in our sample) have $H\delta_A<3\mathrm{\AA}$, the typical threshold to define a spectrum as quiescent.

\subsection{Distributions of $t_Q$ and $\mu_{1.5}$}\label{sec:dists}

\begin{figure*}[!htp]
\includegraphics[width=\textwidth]{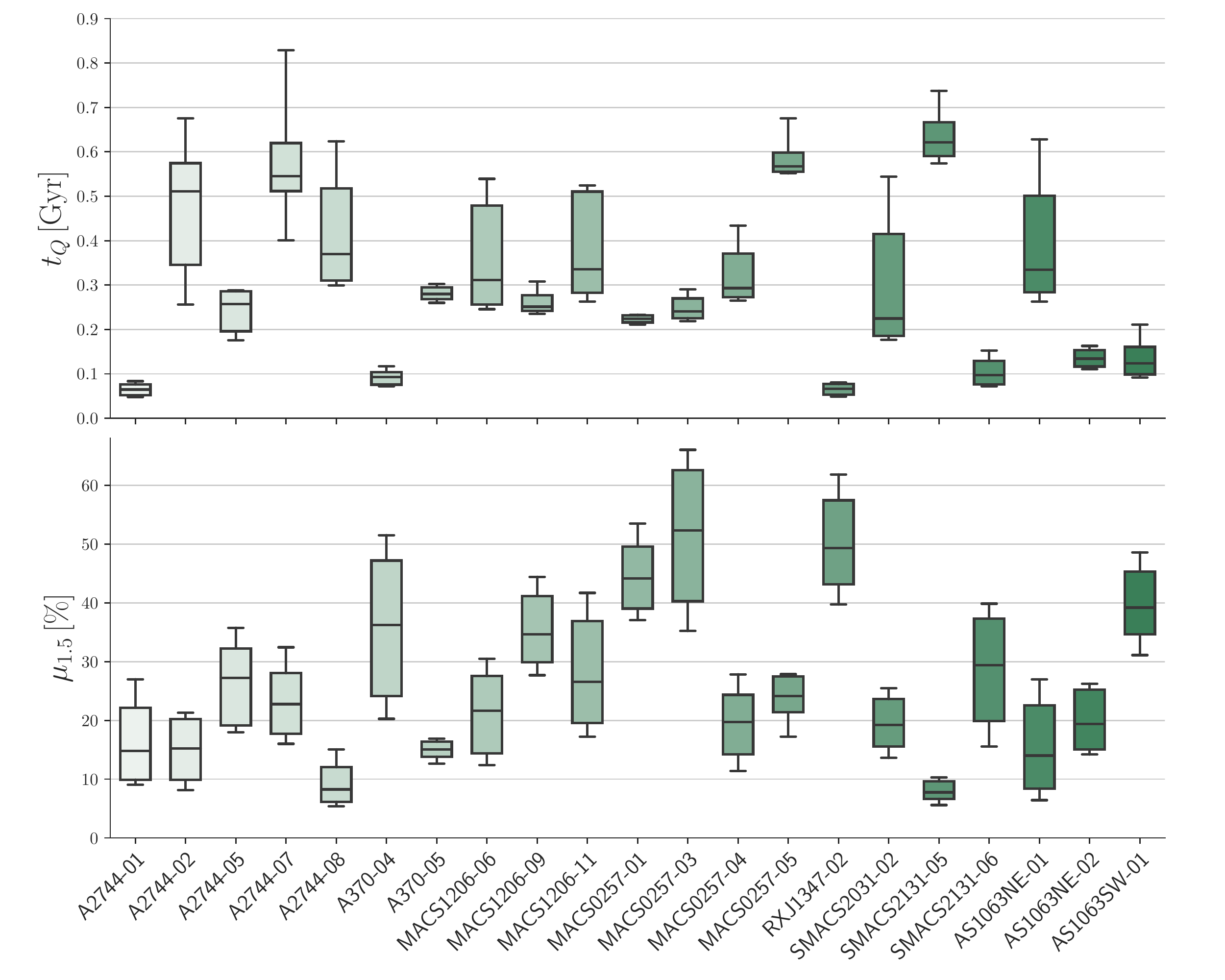}
 \caption{Box plots tracing the distributions of time since quenching ($t_Q$, top) and the fraction of the mass assembled in the past 1.5 Gyrs ($\mu_{1.5}$, bottom). Boxes in shades of green indicate the regions between the 25th and 75th percentiles (interquartile regions), horizontal lines within the boxes indicate the median and whiskers (dark lines outside the boxes) indicate the extent between the 15th and 85th percentiles. The color scale is only to improve readability and does not indicate any physical quantity.}
\label{fig:box}
\end{figure*}

% As explained in section \ref{sec:definitions}, we have reduced the dimensionality of our star-formation histories by calculating two quantities of interest: $t_Q$ and $\mu_{1.5}$ (see section \ref{sec:definitions} for definitions).
In Fig. \ref{fig:box}, we show box plots tracing the distributions of $t_Q$ (top panel) and $\mu_{1.5}$ (bottom panel).
Boxes indicate the interquartile regions with horizontal lines within the boxes showing the median value. Dark lines outside the boxes extend from the 15th to the 85th percentile.

The median values of $t_Q$ vary from 65 Myr for A2744-01 to 621 Myr for SMACS2131-05, with some spaxels reaching times older than 800 Myr. 
The spaxel-by-spaxel mean value for the entire sample is 327 Myr with a standard deviation of 250 Myr.

As for $\mu_{1.5}$, median values go from 7.7\% for SMACS2131-05 to 52.3\% for MACS0257-03, with some spaxel-by-spaxel values reaching more then 60\%. Keeping in mind that 1.5 Gyr corresponds only to $15-17$\% of the time over which these galaxies could form stars (from the Big Bang to the age of the universe at the redshift of observation) and considering that generally the SFR declines with time in normal galaxies, it is noteworthy that the majority of our galaxies have median $\mu_{1.5}$ values greater than 15\%.
The average $\mu_{1.5}$ over all spaxels is 26.10\% with standard deviation of 14.15\%. 
We note that the median values obtained for $\mu_{1.5}$ are generally in the same range as previous works based on integrated spectra \citep[e.g.][]{Pawlik2018, Wild2020}, which is noteworthy considering the different population synthesis methods and post-starburst selection criteria.

\subsection{Quenching directions}\label{sec:maps}

\begin{figure*}[ht]
\includegraphics[width=\textwidth]{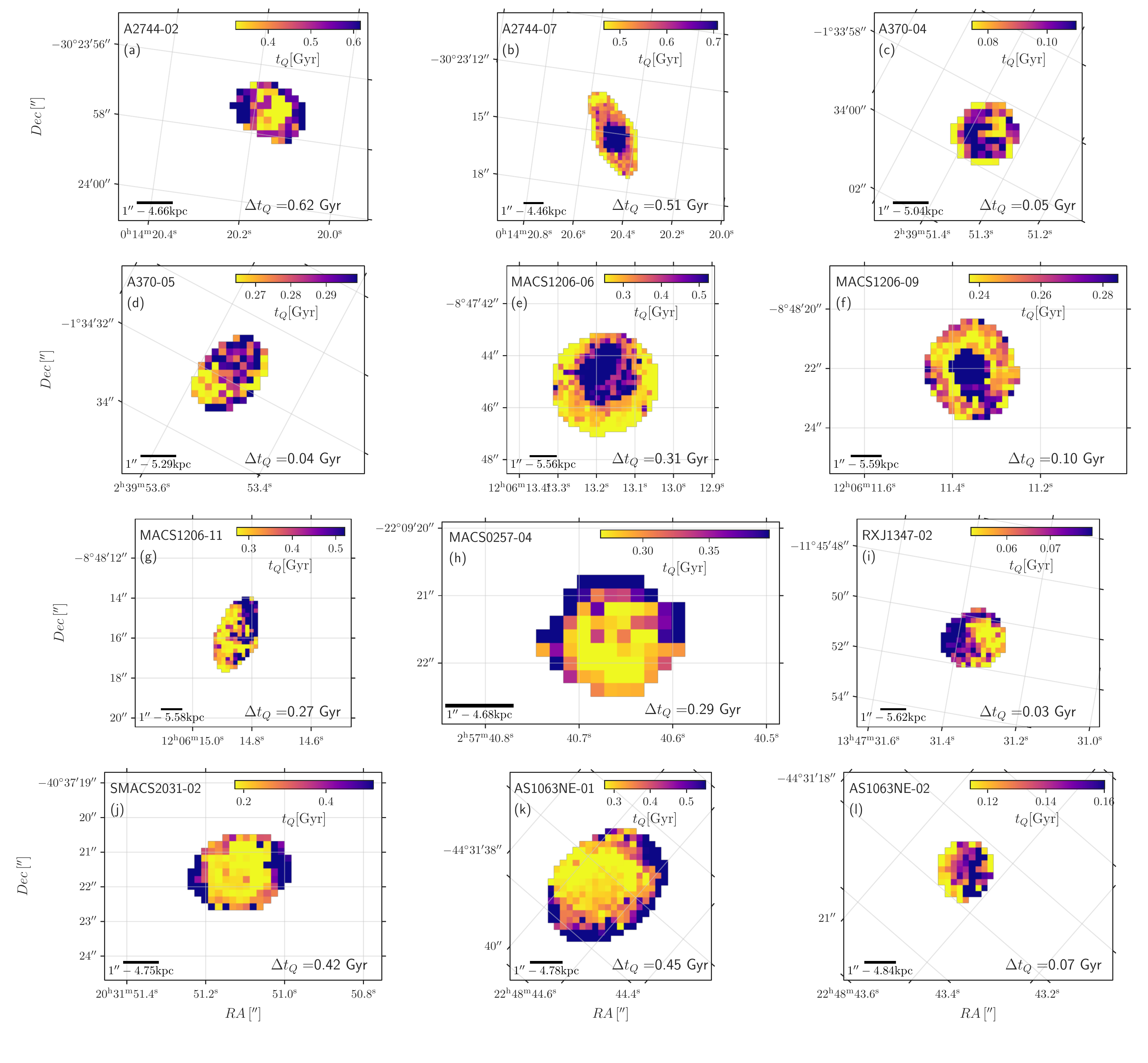}
   \caption{Maps of quenching time ($t_Q$) for all cubes with more then 70 valid spaxels and no foreground objects. Darker colors indicate spaxels quenched at older lookback times.
   Minimum and maximum values of the colorbars are set to the 20\% and 80\% percentiles of each $t_Q$ map. $\Delta t_Q$ values are annotated in each panel.}
\label{fig:tq}
\end{figure*}

We now proceed to a spatially resolved analysis of the SINOPSIS results. In Fig. \ref{fig:tq} we show maps of $t_Q$ for the 12 biggest (in apparent size) galaxies in our sample, for which the datacubes have more then 70 valid spaxels (see definition in section \ref{sec:procedure}). Darker colors indicate older $t_Q$, and  the limits of the colorbars are set to the 20\% and 80\% percentiles of $t_Q$ in each galaxy. 
The variation of $t_Q$ within each galaxy ($\Delta t_Q$) is annotated in each panel, this parameter is an estimate of how quickly the quenching happened. To ensure a robust estimation and prevent eventual outlier spaxels from biasing our results, we define $\Delta t_Q$ as the difference between the 90th and 10th percentiles of $t_Q$, instead of the difference between the maximum and minimum values.

\begin{figure*}[ht]
\includegraphics[width=\textwidth]{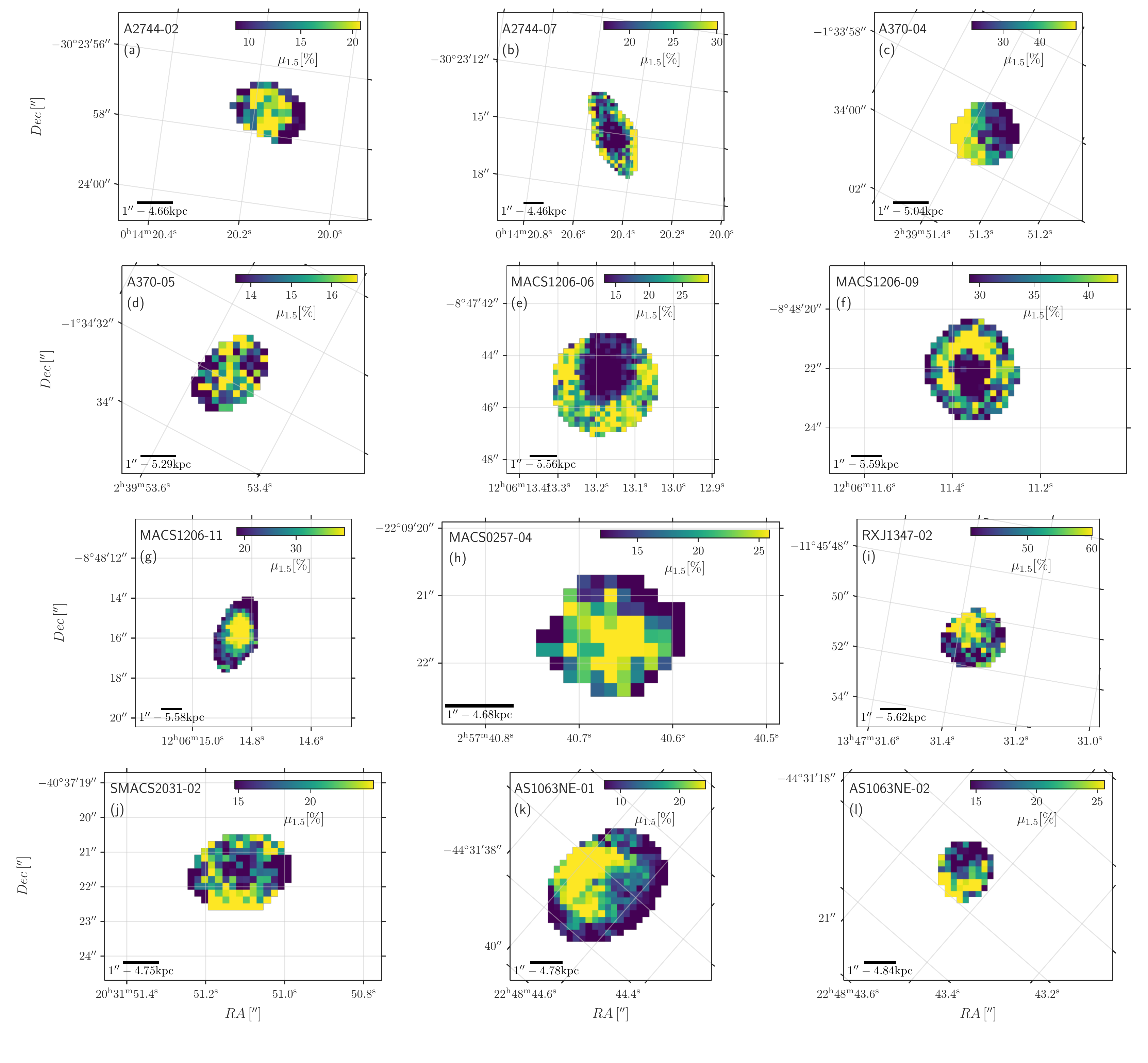}
   \caption{Maps of the fraction of mass assembled in the past 1.5 Gyrs ($\mu_{1.5}$) for all cubes with more then 70 valid spaxels and no foreground objects. Minimum and maximum values of the colorbars are set to the 20\% and 80\% percentiles of each map.}
\label{fig:mu15}
\end{figure*}

The maps in Fig. \ref{fig:tq} indicate a diversity of quenching patterns. 
For 4 out of 12 galaxies the quenching clearly happened first in the outskirts, this is the case for A2744-02, MACS0257-04, SMACS2031-02 and AS1063NE-01 (panels a, h, j and k). This outside-in pattern is reminiscent of the truncated gas disks observed in galaxies undergoing ram-pressure stripping \citep[e.g][]{Marco2017, Poggianti2019JW100}. 
We note that \cite{Deugenio2020a} also finds centrally concentrated starbursts in post-starbursts in environments that are not prone to ram-pressure, where this stellar population gradient is attributed to mergers. However, as we will shown in Fig. \ref{fig:vstar}, the stellar kinematics maps of our galaxies are very regular and inconsistent with the merger hypothesis.

For A370-04, A370-05, MACS1206-11, RXJ1347-02 and AS1063NE-02 (panels c, d, g, i and l) the $t_Q$ distribution is less clear, following a side-to-side or irregular pattern that is also consistent with ram-pressure stripping. The low dynamic range of $t_Q$ in these galaxies (except for MACS1206-11 in panel g) points to a very fast quenching process, uniform over the whole galaxy.

For 3 galaxies (A2744-07, MACS1206-06 and MACS1206-09, panels b, e and f), the $t_Q$ maps indicate an inside-out quenching, which is consistent with the stellar population gradients of the general population of star-forming galaxies (older in the center). 
In the cases of A2744-07 and MACS1206-06, we find indication of AGN emission in the central region of the stellar disks (see section \ref{sec:sample}).

For MACS1206-09, despite the general inside-out pattern we do see that the outer parts of the disk quenched slightly earlier, pointing to a combination of inside-out and outside-in. This pattern could be interpreted as the combination of inside-out quenching due to AGN and outside-in quenching due to ram-pressure, as found by \cite{George2019}. 
We note that this pattern should not be interpreted as peculiar, as there is evidence in the literature for an enhanced AGN fraction among galaxies undergoing ram-pressure stripping \citep[e.g][]{Poggianti2017, Ricarte2020, Peluso2021}, although this is still debated \citep[see][]{Roman-Oliveira2019, Boselli2021}
Despite this duality, we choose to define this galaxy as quenched from the inside out as this is the most conspicuous pattern.

The maps of $\mu_{1.5}$ for the same 12 galaxies are shown in Fig. \ref{fig:mu15}.
To improve visualization we saturate the color scales in the 20th and 80th percentiles of $\mu_{1.5}$ in each galaxy.
In most cases, the $\mu_{1.5}$ maps follow trends that are similar to what we see for $t_Q$.
For the galaxies that show an outside-in pattern in the $t_Q$ maps (A2744-02, MACS0257-04, SMACS2031-02 and AS1063NE-01), the same trend is clearly reproduced for 3/4. In the case of SMACS2031-02 (panel j), the $\mu_{1.5}$ morphology is slightly more complex, as the central region of this galaxy has low mass fractions. 
This can be understood 
%as 
the central regions of the galaxy 
%being 
were more gas-poor prior to the quenching event and thus not assembling large mass fractions during the burst.
On the other hand, MACS1206-11 shows a clear outside-in pattern in the $\mu_{1.5}$ map, while the $t_Q$ map is less clear, showing something between a side-to-side and an outside-in pattern. In this case, we understand that it is safer to classify the galaxy as quenched from the outside-in, since $\mu_{1.5}$ is a more robust parameter and the $\Delta t_Q$ of this galaxy is larger then the typical value associated with side-to-side/irregular quenching patterns.

\subsection{Putting the pieces together}\label{sec:maptrends}

\begin{figure}[ht!]
\includegraphics[width=\columnwidth]{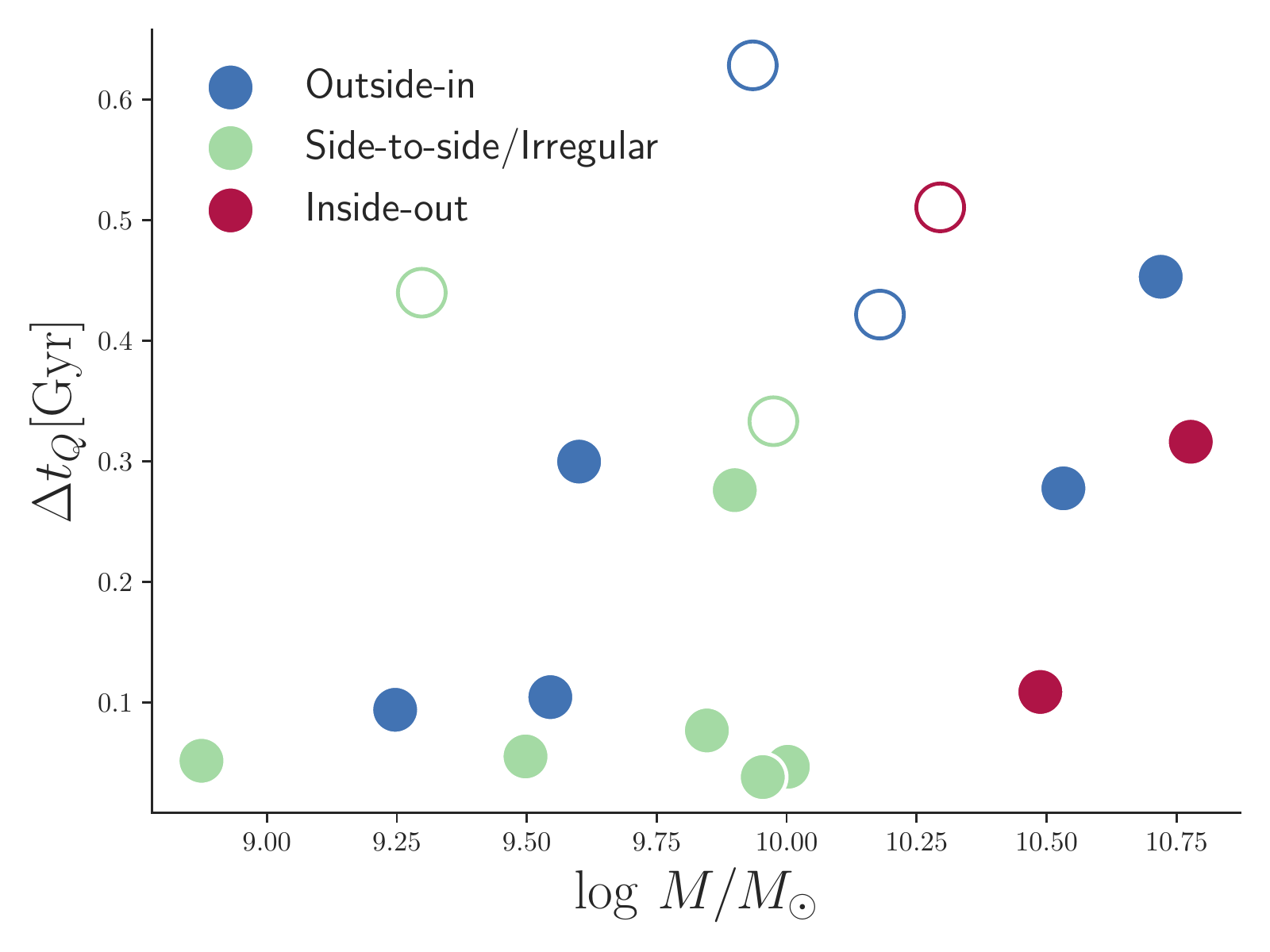}
   \caption{Log stellar mass ($\log\,M/M_\odot$) versus the variation of $t_Q$ within each galaxy ($\Delta t_Q$). Colors indicate the quenching direction measured from the $\mu_{1.5}$ and $t_Q$ maps. Galaxies without a clear enhancement of SFR prior to quenching are shown as open circles.}
\label{fig:mdeltat}
\end{figure}

So far, we have presented results in a case-by-case basis, which is necessary due to the diversity of quenching histories in our sample. However, within this diversity lies a general trend.
In Fig. \ref{fig:mdeltat}, we plot log stellar mass against $\Delta t_Q$ coloring galaxies according to the direction of their quenching maps. Besides the galaxies already shown in Fig. \ref{fig:tq} and Fig. \ref{fig:mu15}, we also include classifications for other 5 galaxies that are not among the biggest (fewer then 70 valid spaxels) but still display clear patterns: these are A2744-01 MACS0257-01, MACS0257-03, MACS0257-05, and SMACS2131-05. We still omit galaxies where part of the stellar disk is obscured by a foreground object, which can affect the observed quenching patterns and render them unclear.

Fig. \ref{fig:mdeltat} shows that (i) all of the side-to-side/irregular quenching patterns are in low-mass galaxies ($\log M/M_\sun<10.5$) and (ii) all of the inside-out patterns are more massive galaxies.
For the low-mass end, we observe generally lower $\Delta t_Q$ (fast quenching) with either a side-to-side pattern or an irregular/unclear morphology in the $t_Q$ and $\mu_{1.5}$ maps. Our interpretation, based on the classical \cite{Gunn&Gott} model, is that the anchoring force that binds the gas to the stellar disk is generally lower, making it easier for ram-pressure to quickly remove the gas.
In more massive galaxies we expect AGN to be more common \citep{Kauffmann2003, Lopes2017, Sanchez2018}, and stellar population gradients to be more strongly affected the evolution of the galaxy before entering the cluster. Indeed, for high-mass galaxies we observe some inside-out patterns left by their previous evolution in a low density environment and/or AGN-induced quenching. Note that this does not mean that the effects of ram-pressure should be completely discarded for these galaxies, just that their stellar population gradients are more strongly shaped by other processes. In fact, as stated before, for MACS1206-09 the maps point to a combination of secular growth and ram-pressure, as both the inner regions and the outskirts have older $t_Q$ and lower $\mu_{1.5}$.

The stellar mass range is unconstrained for galaxies displaying outside-in quenching patterns. These seem to be objects that, while affected by ram-pressure, are able to better retain their gas, leading to a
longer global timescale for stripping of the gas (longer $\Delta t_Q$). Since the stellar mass surface density is lower in the outskirts, these regions are stripped first and the starburst is centrally concentrated, ultimately leading to the outside-in gradients that we observe.

Galaxies lacking a clear enhancement of SFR prior to quenching (as seen in their integrated SFHs, Fig. \ref{fig:intsfh}) are shown as open circles in Fig. \ref{fig:mdeltat}. These objects have typically large $\Delta t_Q$ ($>300$ Myrs) regardless of the quenching direction. These longer timescales can be interpreted as indication that the lack of a burst is not related to how much gas was initially available for star-formation, but to other factors (e.g more circular orbits within the cluster, feedback, etc) that lead to the slow (when compared to other post-starbursts) conversion of this gas into stars.

In addition to what was previously discussed, Fig. \ref{fig:mdeltat} also shows that there is a certain relation between $\log\,M/M_\odot$ and $\Delta t_Q$. In particular, it is interesting to see this relation for galaxies displaying an outside-in quenching pattern: although these are observed in a wide range of masses, the quenching seems to happen faster at lower masses.
We note that, although our sample is large enough so we can identify a variety of objects, it is too small to allow us to reliably study statistical trends, and larger samples will be required to confirm this interpretation.

\bigskip

\section{Special cases}\label{sec:special}

Some objects in our sample have particularly interesting features revealed by the analysis of their emission line maps. In this section, we deviate from the main thread of the paper to concentrate on those cases. In section \ref{sec:tails} we showcase two objects with tails of ionized gas, and in section \ref{sec:A2744-07} we take a closer look at the emission line properties of A2744-07.

\subsection{Tails of ionized gas}\label{sec:tails}

\begin{figure}[ht]
\includegraphics[width=\columnwidth]{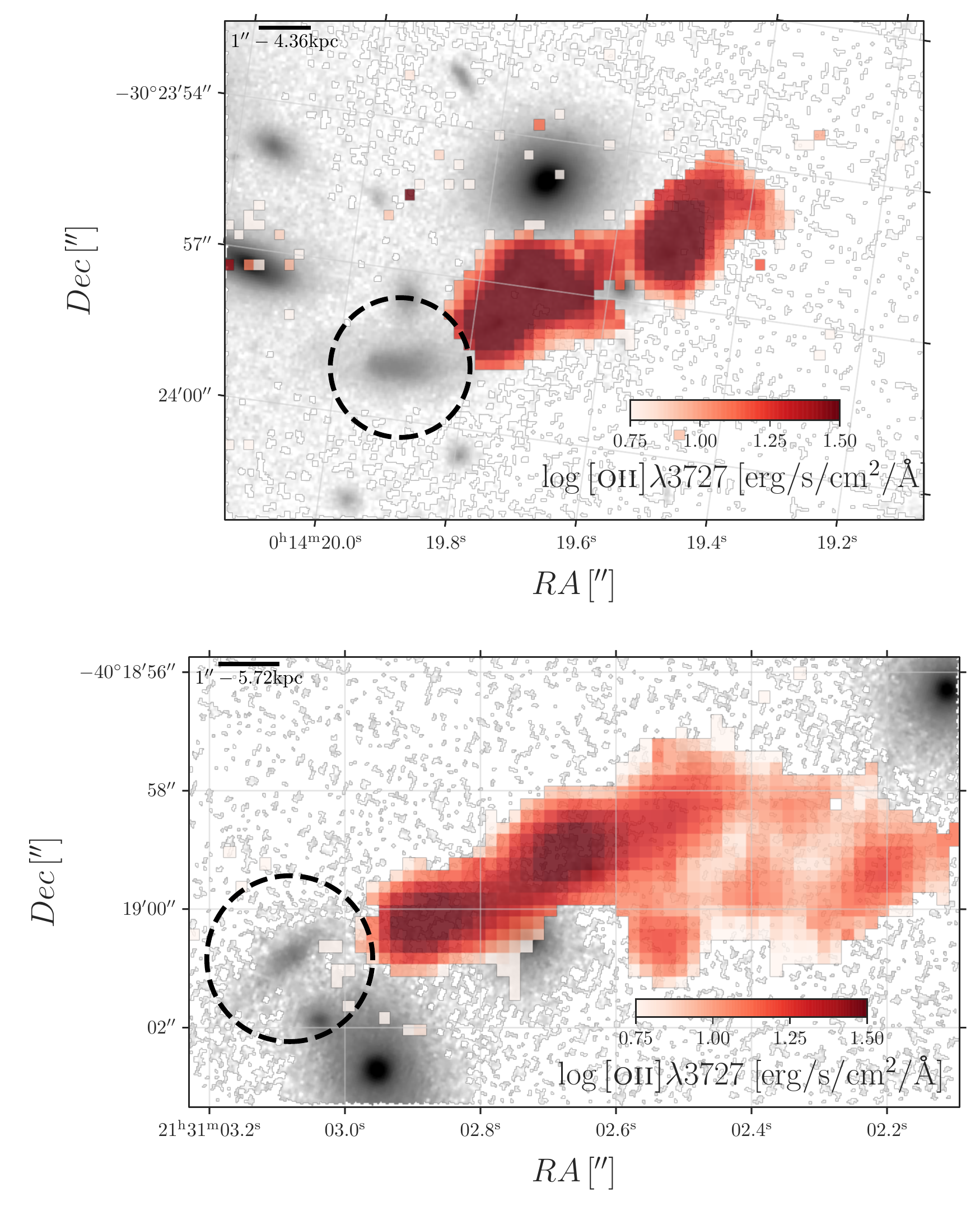}
   \caption{HST F814W images (gray) of A2744-01 (top) and SMACS2131-06 (bottom) with $\Oii$ flux (in $\mathrm{erg/s/cm^2}$) measured from the MUSE data cubes shown in red. The position of the galaxies is marked by dashed circles of arbitrary radius centered in the HST coordinates of each galaxy.}
\label{fig:tail_maps}
\end{figure}

We have identified two post-starbursts that, although lacking any emission lines in their stellar disks, display long tails of ionized gas, these are A2744-01 and SMACS2131-06. 
We note that for A2744-01 these tails have already been identified by \cite{Owers2012} and \cite{Moretti2021}.
This intriguing feature is shown in Fig. \ref{fig:tail_maps}, where we plot HST F814W images of these two objects with $\Oii$ flux in red. 
These tails can be interpreted as the smoking gun of ram-pressure stripping, setting these objects as missing links between post-starbursts and jellyfish galaxies.

\begin{figure}[ht]
\includegraphics[width=\columnwidth]{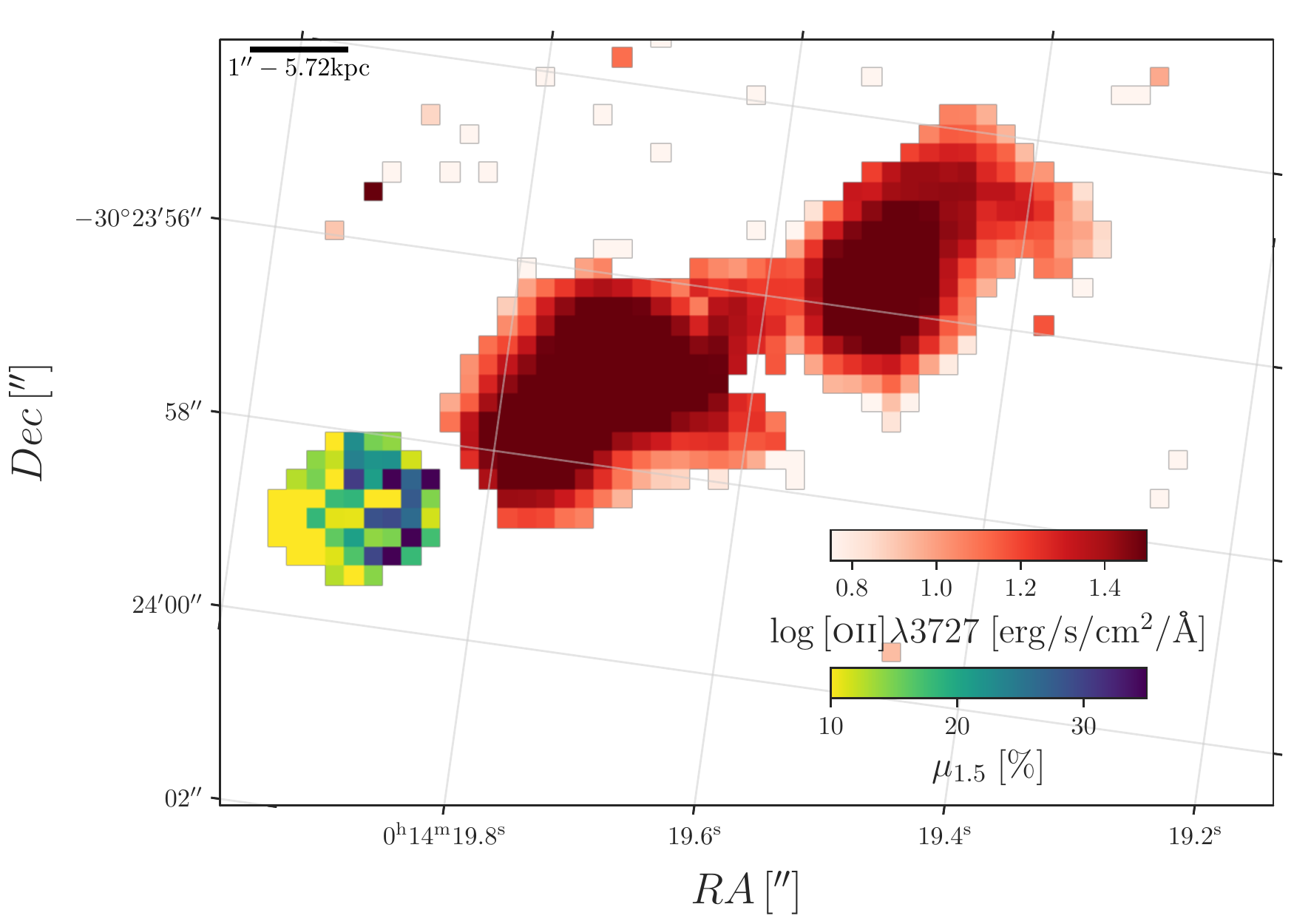}
   \caption{Maps of $\mu_{1.5}$ (yellow to blue color map) and $\Oii$ flux (red color map) for the galaxy A2744-01. }
\label{fig:tail_map_A2744_01}
\end{figure}

The case of A2744-01 is particularly interesting. 
In Fig. \ref{fig:tail_map_A2744_01}, we plot the map of $\mu_{1.5}$ for this galaxy (yellow to blue color map) and the flux map in of the $\Oii$ line (red color map). 
The $\mu_{1.5}$ map 
has a side-to-side gradient that is aligned with the tail of ionized gas in such a way that spaxels closer to the tail have larger fractions of young stellar populations (more prominent bursts). 
Unfortunately, we are unable to do the same analysis for SMACS2131-06, as in the MUSE data cube it is strongly blended with a foreground object in the south, which prevents us from reliably probing the stellar population gradient.

Other properties of A2744-01 are also worth mentioning.
Of all galaxies in our sample, it is the one with the lowest mass ($\log M_\star/M_\odot = 8.9$), lowest median $t_Q$ (64 Myr, see Fig. \ref{fig:box}) and a very low $\Delta t_Q$ of 51 Myr. It also has very strong  features with a blue spectrum (see top panel of Fig. \ref{fig:fits}), integrated $H\delta_A$ of 6.78$\mathrm{\AA}$ and a very clear recent burst in the star-formation history (see first panel in Fig. \ref{fig:intsfh}). Its median  $\mu_{1.5}$ is 14.78\%. 
% In particular, the low $t_Q$ of 64 Myr and the stellar population gradient aligned with the tail (with spaxels closer to the tail having more significant recent bursts) show that this galaxy is caught in its transition from jellyfish to post-starburst.

\subsection{The case of A2744-07}\label{sec:A2744-07}

\begin{figure}[ht!]
\includegraphics[width=\columnwidth]{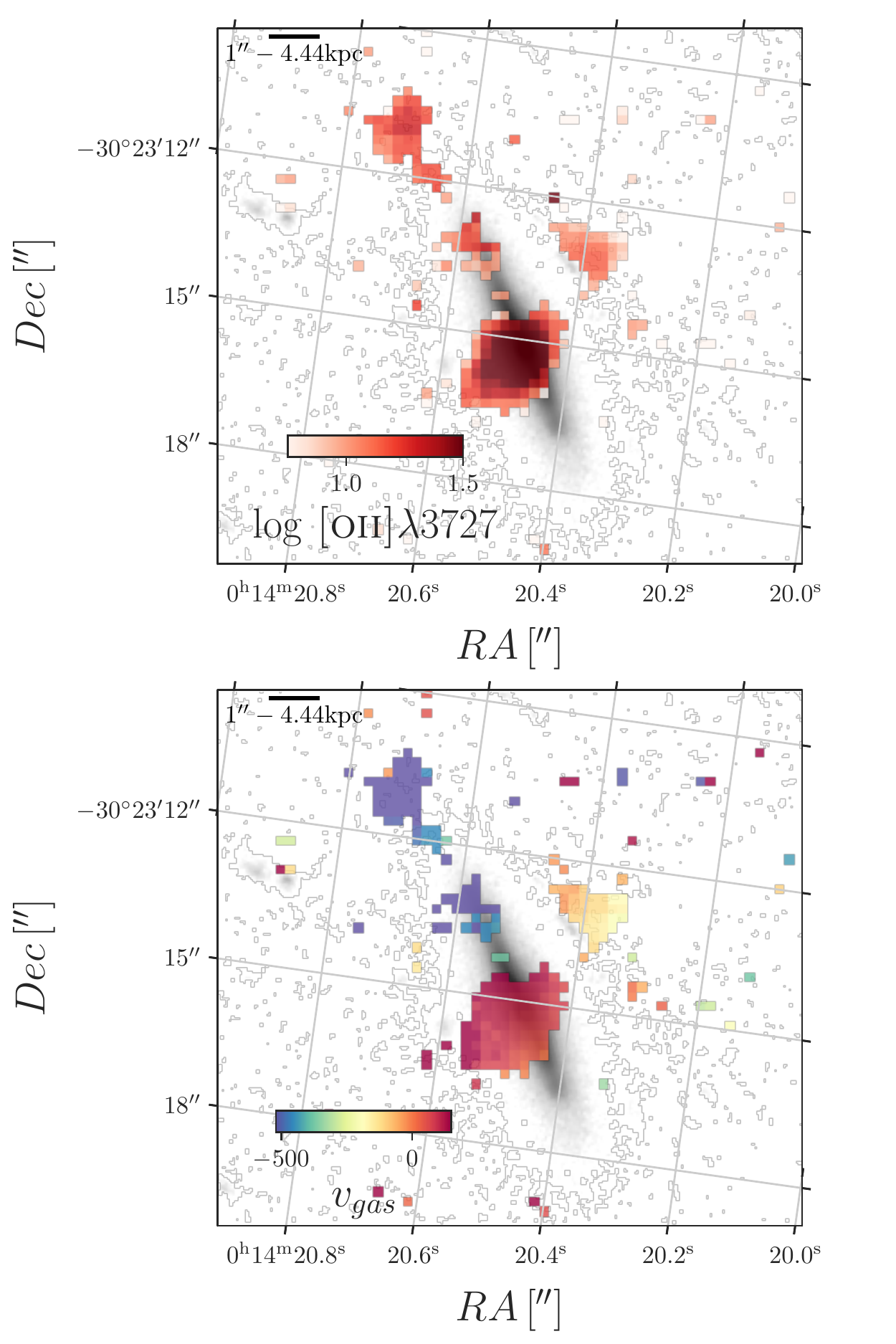}
   \caption{MUSE $\Oii$ map for all spaxels where the line is detected with $S/N>3$ (top panel) and $\Oii$ velocity field (bottom panel) for the galaxy A2744-07. HST F814W image is shown in gray scale in the background.}
\label{fig:oii_vel}
\end{figure}

Another object with peculiar emission line features is A2744-07. In Fig. \ref{fig:oii_vel} we show the $\Oii$ flux (top) and velocity (bottom) for this object.

Besides the AGN in the central region (see section \ref{sec:sample}),  
two strips of gas can be seen in the map, one to the North-East, extending $\sim$ 20 kpc along and beyond the stellar disk, and one to the East, extending $\sim$ 12 kpc perpendicular to the stellar disk. In addition to $\Oii$, these features are also visible in $H\alpha$, but not in $\Oiii$, which points to a 
soft ionization source, although the strips seem to be spatially correlated with the central AGN. 
Both gas strips are blueshifted with respect to the central region (bottom panel of Fig. \ref{fig:oii_vel}), in the case of the northern strip the velocities reach up to $\sim-500\,\mathrm{km/s}$.

We are unable to determine if these features are caused by the AGN, by ram-pressure or (most likely) a complex combination of the two. Given that ram-pressure stripping is known to trigger (or maybe be 
%triggered 
helped by) an AGN \citep[e.g][]{Poggianti2017, Ricarte2020}, we should expect some transition objects such as A2744-07 to display features of both AGN and ram-pressure.

\section{Further clues on ram-pressure stripping}\label{sec:discussion}

Our results so far indicate that ram-pressure stripping is the main driver of fast quenching in the centers of clusters at intermediate redshift. 
This hypothesis is strengthened by several results in the literature that also point to ram-pressure as the most common path to the post-starburst phase in dense environments \citep{Poggianti2004, Poggianti2009, Vulcani2020}.  
In this work, the main evidence for this comes from the direction of the quenching traced by $t_Q$ and $\mu_{1.5}$, as these maps show that most galaxies are quenched from the outside-in or display a side-to-side/irregular stellar population gradient. Both features are consistent with the stripping of gas via ram-pressure.
Inside-out quenching patterns expected from AGN-driven quenching are observed only in 3 galaxies, all of which are in the high-mass end of the distribution for our sample.

\begin{figure}[ht]
\includegraphics[width=\columnwidth]{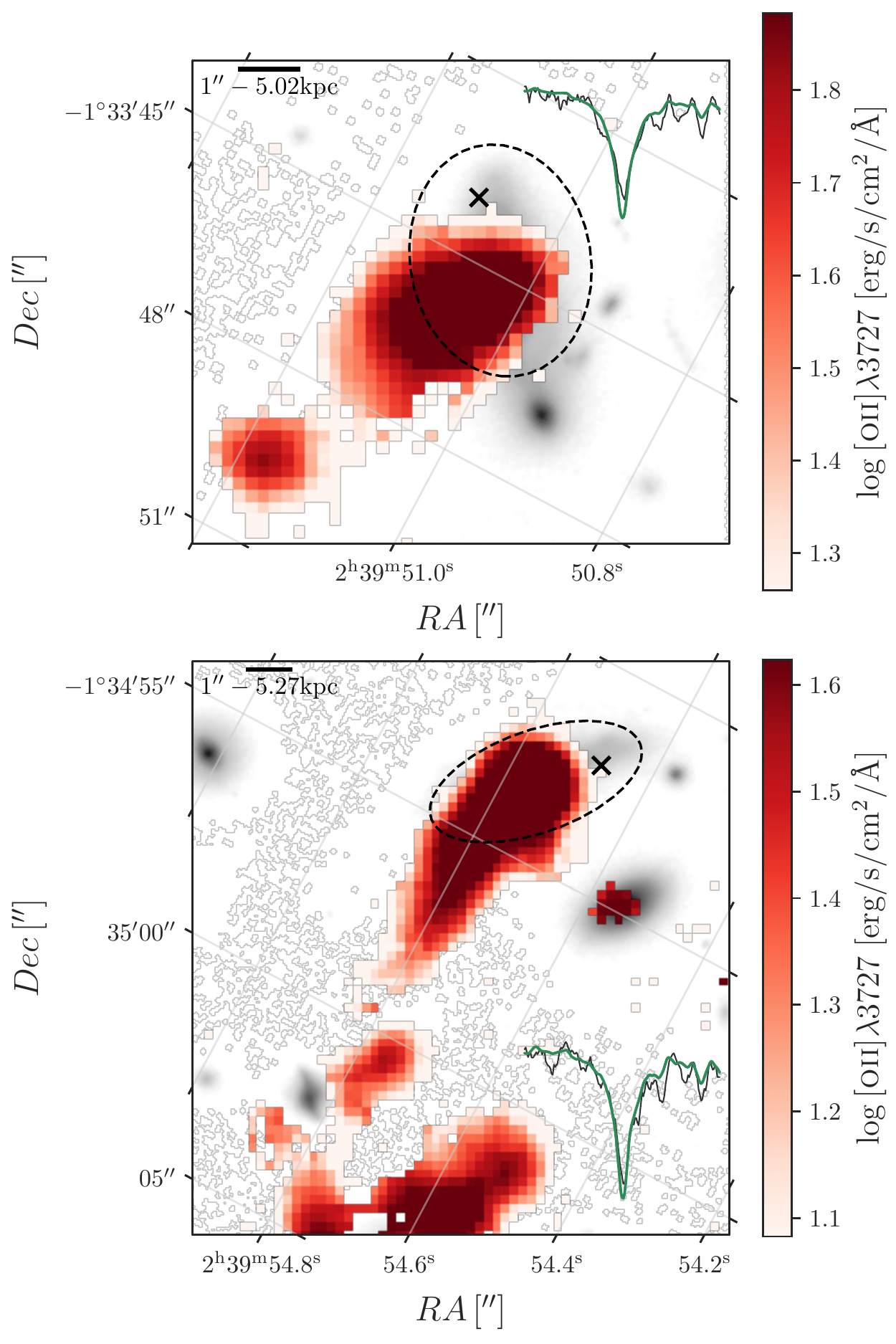}
   \caption{$\Oii$ flux map of two jellyfish galaxies from \cite{Moretti2021}, A370-06 in the top and A370-08 in the bottom, with dashed ellipses showing the extension of the stellar disk. A dark "x" marks the position of a spaxel that has recently been stripped of gas, the model (green) and observed (black) spectra of these spaxels in a 75\AA\ window centered on the $H\beta$ line are plotted on the right-hand side. The HST F814W image is shown in gray scale in the background.}
\label{fig:psb_jelly}
\end{figure}

\begin{figure*}[ht]
\includegraphics[width=\textwidth]{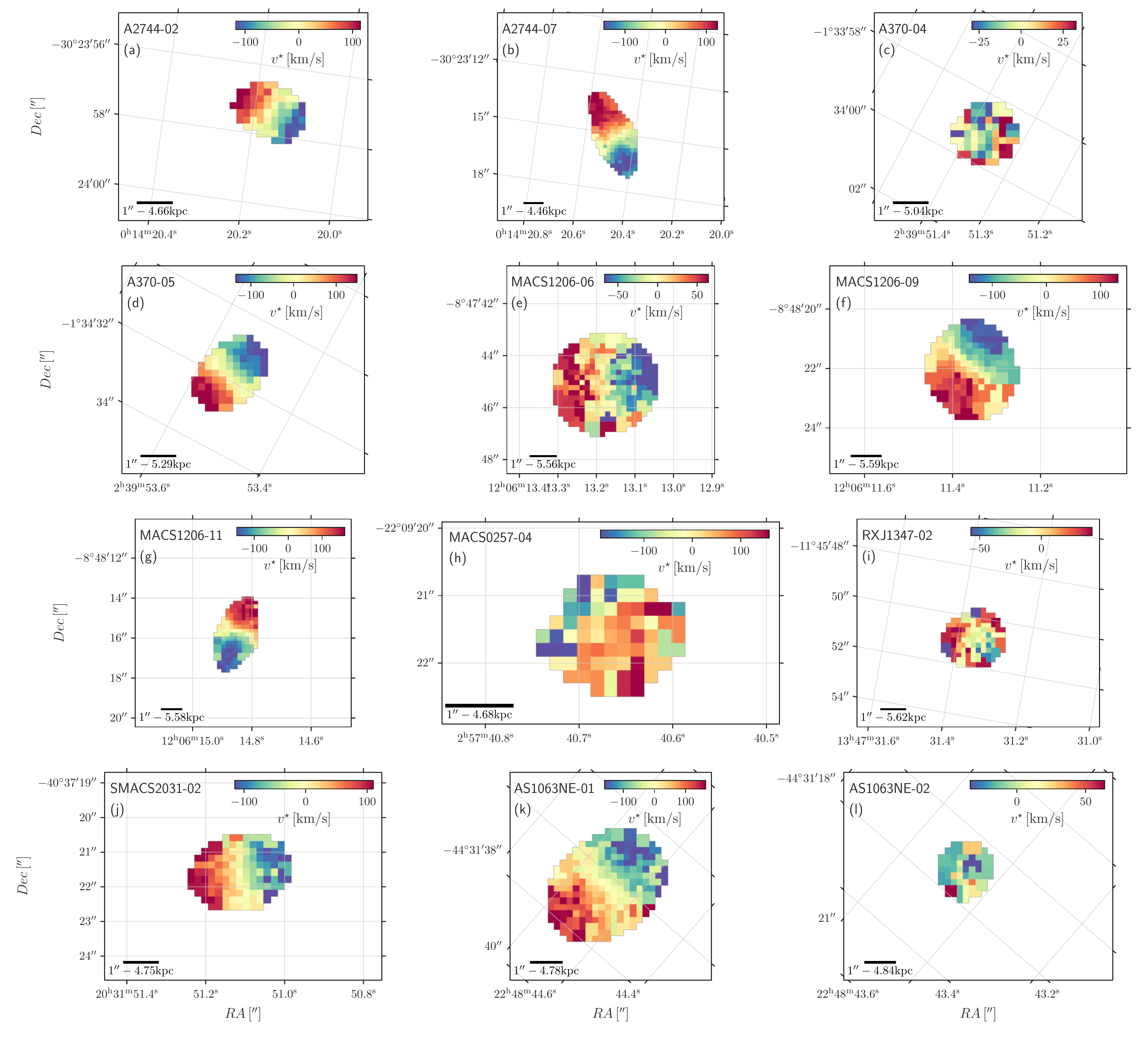}
  \caption{Stellar velocity ($v^\star$) maps derived with \textsc{pPXF} for the same 12 galaxies shown in Figures \ref{fig:tq} and \ref{fig:mu15}.}
\label{fig:vstar}
\end{figure*}

Complementary evidence comes from two special cases (shown in Fig. \ref{fig:tail_maps}) for which we find tails of ionized gas indicating that the gas was very recently stripped.
This connection between post-starburst and jellyfish galaxies can also be explored from the other way around, i.e identifying post-starburst features in galaxies undergoing ram-pressure stripping, as done by \cite{Marco2017} and \cite{Poggianti2019JW100} for jellyfish galaxies in the local universe.
Thus, it is natural to attempt the same analysis for stripped galaxies in the same clusters from which we selected our sample of post-starbursts.
In Fig. \ref{fig:psb_jelly} we show maps of $\Oii$ flux from two jellyfish galaxies from the \cite{Moretti2021} sample, A370-06 on the top panel and A370-08 on the bottom panel. Dashed ellipses in each panel indicate the extent of the stellar disk. For each of these galaxies we choose a spaxel that has already been stripped of gas (marked with a dark "x") and plot the corresponding spectra around the $H\beta$ line.
The spectra in these stripped regions show strong $H\beta$ absorption lines, and the same happens for other Balmer lines, indicating that these objects are likely to be the progenitors of post-starburst galaxies.

A crucial piece of information for confirming ram-pressure stripping is the stellar velocity field of a galaxy \citep[e.g][]{Vulcani2020}. Disturbed stellar kinematics can be associated with gravitational interactions or mergers, while ram-pressure affects only the gas and leaves the stellar kinematics unperturbed.
Unfortunately, at this redshift, due to the spatial resolution and signal to noise ratio it is not possible to establish if an irregular velocity map corresponds to actually disturbed stellar kinematics or just noise. Therefore, irregular kinematic maps should be interpreted as ambiguous. However, it is possible to confirm when the kinematics is globally regular. 

Stellar velocity fields derived with \textsc{pPXF} for the same 12 galaxies of Figures \ref{fig:tq} and \ref{fig:mu15} are shown in Fig. \ref{fig:vstar}. The stellar kinematics is usually very regular. Exceptions are A370-04, AS1063NE-02 and RXJ1347-02, all of which are faint and have small apparent size, for which the velocity maps are not reliable. 
Thus, the kinematical maps for these galaxies are very uncertain. 
We interpret these regular kinematical maps as complementary evidence that ram-pressure is the main driver of fast quenching in our sample.  

\begin{figure}[ht]
\includegraphics[width=\columnwidth]{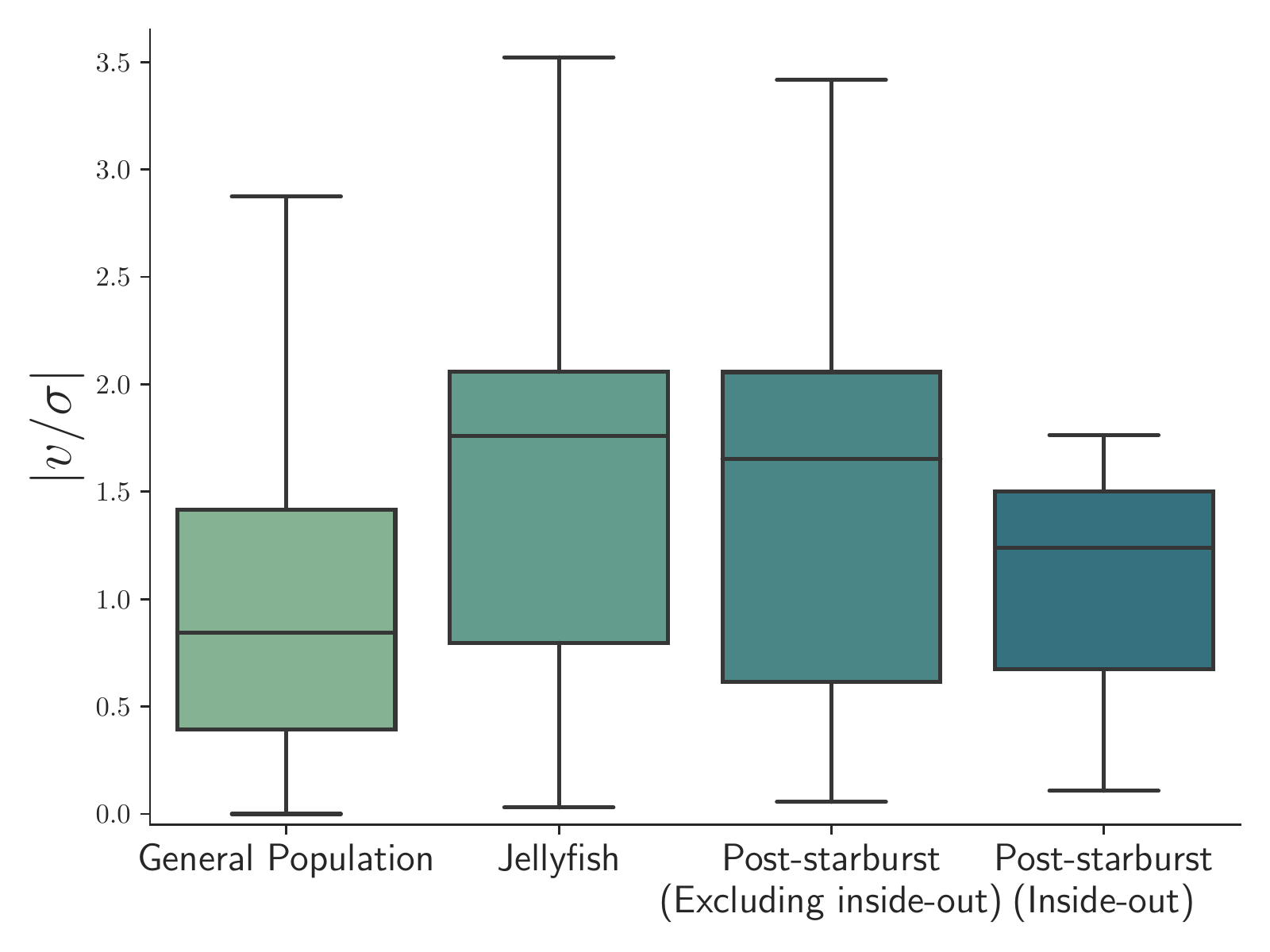}
  \caption{Box plots tracing the absolute value of $v/\sigma$ for (from left to right) the General population of cluster galaxies, Jellyfish galaxies from the same clusters, post-starbursts not including the ones with inside-out quenching, and post-starbursts with inside-out quenching. Boxes between the interquartile regions, horizontal lines within the boxes indicate the median, whiskers extend out to 1.5 times the interquartile regions.}
\label{fig:ps}
\end{figure}

Another relevant piece of the puzzle is the galaxy environment. To probe this, we calculate the absolute value of the peculiar velocity of each galaxy relative to the cluster redshift, normalized by the cluster velocity dispersion ($|v/\sigma|$).
The value of $|v/\sigma|$ gives us an indication to the infall stage of each galaxy. Objects with higher $|v/\sigma|$ are more likely to be rapidly infalling and experiencing ram-pressure stripping, while objects with lower $|v/\sigma|$ are generally more likely to be relaxed.

Box plots tracing the distributions of $|v/\sigma|$ for different galaxy classes are shown in Fig. \ref{fig:ps}. We divide the post-starburst population (marked as PSB in the Figure) in two sub-classes: the ones with outside-in or side-by-side/irregular quenching patterns and the ones with inside-out quenching patterns.
Note that in the case of inside-out quenched post-starbursts the box plot is produced with only 3 galaxies, the $|v/\sigma|$ values for these galaxies correspond to the median and to the extremes of the whiskers in the box plot.

The box plots in Fig. \ref{fig:ps} show that post-starbursts with outside-in or side-to-side/irregular quenching patterns have a $|v/\sigma|$ distribution similar to the one of jellyfish galaxies. 
While post-starbursts with inside-out quenching patterns are closer to the general population of cluster galaxies.
This serves as further evidence that the post-starbursts with outside-in or side-to-side/irregular quenching patterns had their gas removed by ram-pressure stripping, while inside-out quenching patterns 
%are 
may be related also to other processes such as AGN feedback.

\section{Conclusions}\label{sec:conclusions}

In this work, we have presented results from the analysis of MUSE integral field spectroscopy of 21 post-starburst galaxies in the centers of 8 clusters at $z\sim0.3-0.4$.

We used the SINOPSIS spectral synthesis code to retrieve spatially-resolved star-formation histories of all 21 objects, finding a clear enhancement in SFR prior to quenching in 16 cases. 
From these, we calculate the time since quenching ($t_Q$) and the fraction of stellar mass assembled in the past 1.5 Gyr ($\mu_1.5$).

We find that
most galaxies in our sample have quenched their star-formation from the outside-in (7 objects) or show a side-to-side/irregular pattern (8 objects).
Only 3 objects show an inside-out quenching pattern, all of which are at the high-mass end of our sample. For 2 of them the quenching can be associated with an AGN.
For 3 objects the quenching direction is unclear.
 The variation in quenching times within each galaxy ($\Delta t_Q$) correlates (although weakly) with galaxy mass, in the sense that less massive galaxies are more rapidly quenched.

For two objects in our sample, we identify tails of ionized gas showing that these galaxies are caught in a transition from jellyfish to post-starburst. In the case of A2744-01, we are able to identify a stellar population gradient in the direction of the tail: spaxels closer to the tail have assembled higher fractions of stellar mass in the recent past.

We have shown that the stellar kinematics of galaxies in our sample is generally regular. Also, post-starbursts displaying an outside-in or a side-to-side/irregular quenching pattern have a distribution of absolute velocities within the cluster similar to the one of galaxies currently undergoing ram-pressure stripping. On the other hand, the three galaxies for which we identify an inside-out quenching pattern, the velocities within the cluster are 
closer to the ones of the general population of cluster galaxies.

As a whole, our results point to ram-pressure stripping as the main driver of fast quenching in dense environments at intermediate redshift, with active galactic nuclei being relevant only for galaxies of high stellar mass ($\log \, M/M_\sun > 10.5$).

\section*{Acknowledgments}
AW thanks Catarina Aydar and Daniel Ruschel Dutra for important discussions about this work, and Abilio Mateus for providing a first contact with post-starburst galaxies.
We thank the anonymous referee for suggestions that have improved the paper.
This project has received funding from the European Research Council (ERC) under the European Union's Horizon 2020 research and innovation program (grant agreement No. 833824, GASP project). 
B.~V. and M.~G. also acknowledge the Italian PRIN-Miur 2017 (PI A. Cimatti).
We acknowledge funding from the INAF main-stream funding programme (PI B. Vulcani).
J.F. acknowledges financial support from the UNAM- DGAPA-PAPIIT IN111620 grant, México.
GB acknowledges financial support from the National Autonomous University of M\'exico (UNAM) through grant DGAPA/PAPIIT BG100622.

\bibliographystyle{aasjournal}
\bibliography{references}

\end{document}